\documentclass[preprint,authoryear,3p,times]{elsarticle}

\usepackage[latin1]{inputenc}

\usepackage{graphicx}
\usepackage{subfigure}
\usepackage{bm}
\usepackage{hyperref}

\usepackage{empheq}
\usepackage{xspace}
\usepackage{color}
\usepackage{amssymb}
\usepackage{bbold}
\usepackage{booktabs}
\usepackage{appendix}
\usepackage{placeins}

\hypersetup{
bookmarksnumbered = true,
unicode=false,          
pdftoolbar=true,        
pdfmenubar=true,        
pdffitwindow=false,     
pdfstartview={Fit},    
pdftitle={Journal of the Mechanics and Physics of Solids},    
pdfauthor={P.Recho},     
pdfsubject={Subject},   
pdfcreator={Creator},   
pdfproducer={Producer}, 
pdfkeywords={keyword1} {key2} {key3}, 
pdfnewwindow=true,      
colorlinks=true,       
linkcolor=black,          
citecolor=black,        
filecolor=black,      
urlcolor=black,           
pdfdisplaydoctitle = true
}

\usepackage[normalem]{ulem}
\usepackage{color}

\newcommand{\tr}[1]{\textcolor{black}{#1}}

\journal{Journal of the Mechanics and Physics of Solids}

\begin{document} 
\text

\title{Chemo-mechanical model of a cell as a stochastic active gel}

\author[cued]{V. Deshpande}
\ead{vsd20@cam.ac.uk}
\address[cued]{Engineering Department, University of Cambridge, Cambridge, UK}
\author[sissa,pisa]{A. DeSimone}
\ead{desimone@sissa.it}
\address[sissa]{Scuola Internazionale Superiore di Studi Avanzati, I-34136 Trieste, Italy}
\address[pisa]{The BioRobotics Institute and Department of Excellence in Robotics and A.I., Sant'Anna School for Advanced Studies, I-56127 Pisa, Italy}
\author[mems,seua]{R. McMeeking}
\ead{rmcm@engineering.ucsb.edu}
\address[mems]{Mechanical Engineering and Materials Departments, University of California, Santa Barbara, USA}
\address[seua]{School of Engineering, University of Aberdeen, Aberdeen, UK}
\author[liphy]{P. Recho}
\ead{pierre.recho@univ-grenoble-alpes.fr}
\address[liphy]{Universit\'e Grenoble Alpes, Laboratoire Interdisciplinaire de Physique, CNRS, F-38000 Grenoble, France}

\begin{abstract}
While it is commonly observed that the shape dynamics of mammalian cells can undergo large random fluctuations, theoretical models aiming at capturing cell mechanics \tr{often focus on the deterministic part of the motion.} In this paper, we present a  framework that couples an active gel model of the cell mechanical scaffold with the complex cell metabolic system stochastically delivering the chemical energy needed to sustain an active stress in the scaffold. Our closure assumption setting the magnitude of the fluctuations is that the chemo-mechanical free energy of the cell is controlled at a target homeostatic value. Our model rationalizes the experimental observation that the cell shape fluctuations depend on the mechanical environment that constraints the cell. We apply our framework to the simple case of a cell migrating on a one dimensional track to successfully capture the different regimes of the cell mean square displacement along the track as well as the magnitude of the long time scale effective diffusive motion of the cell.

\end{abstract}

\maketitle

\section{Introduction}

The mechanical behavior of an eukaryotic cell is largely controlled by its cytoskeleton, a polymer structure that is fundamentally out of thermodynamic equilibrium. Cytoskeletal processes such as the polymerization turnover of its scaffold and stress-fibre contractility, driven by molecular motors power-strokes, are fuelled by nutrient exchanges of the cell with its environment. Free-energy is released by the hydrolysis of Adenosine Triphosphate (ATP), ultimately serving as the common free-energy source to drive all the mechanically active processes in the cell.

There now exists a very active branch of study broadly known as ``active gel physics'' that aims to develop a continuum medium deterministic formalism for the transduction of the chemical energy within cells to the mechanics of its cytoskeleton \tr{\citep{prost2015active}}. The basic idea behind this approach is to perturb the state of the system (defined subsequently) in the vicinity of an assumed equilibrium state. This allows the use of the formalism of linear out-of-equilibrium thermodynamics and the Onsager symmetry relations.  Active gel theories of this type typically involve dissipative processes such as viscosity, diffusion, chemical reactions and, importantly, a generalised force that drives the system out of equilibrium. Given that the hydrolysis of ATP is the free-energy source of the molecular motors actuating the cytoskeleton, the natural choice for this generalised force $\Delta \mu$ is the difference in the chemical potentials of ATP and of the product of hydrolysis. In the current active gel theories, this chemical affinity is maintained at a fixed value as an intrinsic property of the cell, although there is strong evidence in terms of  cytoskeleton polymerisation \citep{solon2007fibroblast,engler2006matrix} and more direct recent measurements \citep{park2020mechanical} which suggest that the cellular metabolism is modified by the mechanical environment of the cell. Of note, the linear Onsager framework is theoretically valid for a system near equilibrium. In the context of active gels, this would imply that $\Delta \mu$ is small compared to $k_BT$, where $k_B$ is the Boltzmann constant and $T$ the absolute temperature. Under physiological conditions $\Delta \mu$ can be on the order of $20 k_BT$, and the use of the Onsager framework should therefore be regarded as a general guideline in which specific phenomenological non-linearities between generalised forces and fluxes should be introduced if necessary.

\begin{figure}[h!]
\centering
\includegraphics[width=0.5\textwidth]{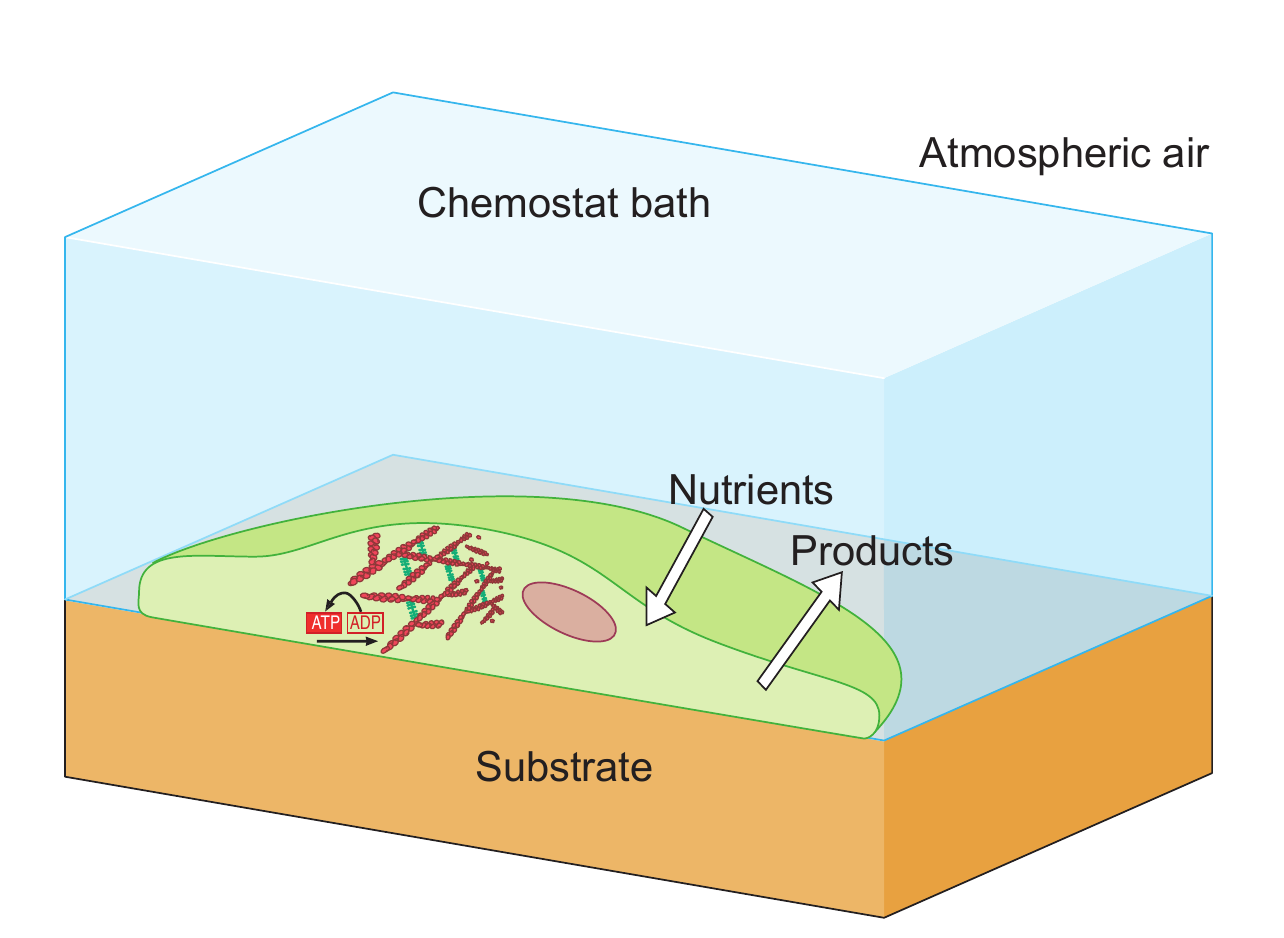}
\caption{\label{fig:scheme_model} Chemo-mechanical model of the cell. The cytoskeleton filaments are represented in red and cross-linked by molecular motors inflicting active stress in the meshwork. The motors are running by using the chemical energy released during the ATP hydrolysis to ADP. The ADP molecules are recycled into ATP thanks to the cell metabolism that consumes nutrients and expels products. The concentration of nutrients and products in the culture medium is maintained to constant values (chemostat).}
\end{figure}

In order to understand some of the assumptions implicit within the active gel theories, consider a typical in-vitro experiment comprising a cell on a substrate immersed in a nutrient-rich medium (nutrient bath); see Fig.~\ref{fig:scheme_model}. While the cell exchanges nutrients (e.g. glucose) and other species (e.g. gases, ions, morphogens) with the surrounding bath, the entire setup is exposed to an external atmosphere and maintained at  constant temperature and pressure. The nutrient bath is assumed to be infinite in extent such that it acts as a \emph{chemostat} maintaining the concentrations of nutrients to a constant level. Within the cell, the ATP hydrolysis reaction fueling the molecular motors in the cytoskeleton is happening spontaneously. \tr{If the cell were a closed system, the extent of this reaction  should relax to equilibrium}. However, a complex and highly regulated system, the cellular metabolism, uses nutrients from the bath to constantly recycle the products of hydrolysis back into ATP. In classical active gel theories, the thermodynamic system considered comprises only the cystoskeleton and it is assumed that the metabolic recycling  spontaneously adjusts to maintain a constant value of $\Delta \mu$. In this case, the role of metabolism is therefore analogous to a Maxwell daemon constantly delivering ATP to the cytoskeleton and removing ADP when necessary, such that the extent of ATP hydrolysis keeps monotonically increasing. The free energy of the thermodynamic system is therefore bottomless \citep{recho2014optimality}. As such, this closure assumption creates no theoretical issues other than the fact that 
an alternative model based on an open system comprising the cell along with its metabolic system that uses the chemical energy in the bath to dynamically allocate the energy resources to the molecular motors might be more insightful.

Our practical motivation to extend the deterministic active gel formalism to incorporate the cell metabolic system stems from the presence of a so-called ``biological noise'' in the cell dynamics.  The fluctuations associated with this noise provide the mechanism for numerous critical biological functions such as the long time-scale spatial exploration of the environment via Brownian-like motility \citep{stokes1991migration, li2008persistent} and the nematic ordering of cells which plays a important role in tissue morphogenesis \citep{gupta2015adaptive}. Biological noise manifests itself through non-thermal fluctuations of cytoskeletal filaments \citep{brangwynne2007force,nadrowski2004active} which then propagate into fluctuations of cellular level observables including cell shape and cell stresses as typically quantified by its tractions on the substrate  \citep{engler2006matrix, solon2007fibroblast}. These fluctuations are usually measured  in terms of the  observables' standard errors which are typically quite large. For example, the variability of the aspect ratio of cells seeded on stiff substrates is on the order of the mean aspect ratio \citep{prager2011fibroblast}. Moreover, unlike inorganic systems where these errors are associated with observation or other experimental errors, variability in the observations for cells is associated with their intrinsically fluctuating response and depend on the cell chemical and mechanical environment: typically, the variability in the observations decreases with decreasing stiffness of the cell environment \citep{engler2006matrix,prager2011fibroblast}. 

There have been numerous attempts to include the effect of biological \tr{fluctuations} in models for cell dynamics. \tr{A lot of them} have borrowed the notion of an effective temperature at a long timescale used to describe the motion of grains in granular media \citep{edwards1989theory} and  extract the effective temperature by either directly fitting to observations \citep{stokes1991migration, fabry2001scaling} or by using physical notions such as persistence length \citep{brangwynne2007force} to then infer an effective temperature. \tr{This notion is clearly not related to its standard  statistical thermodynamics definition involving the average kinetic energy of some internal degrees of freedom but is an effective way to describe the magnitude of some fluctuations occurring at the microscale in a non-equilibrium system.} Such approaches neglect the fact that the biological noise and hence the effective temperature is in fact a function of the cellular environment. \tr{Interestingly, some fully deterministic chemomechanical models of the cell behaviour  can lead to a spatio-temporal chaotic behaviour whose long timescale dynamics can be characterized by an effective diffusion coefficient \citep{dreher2014spiral,stankevicins2020deterministic}. } Recently, \cite{shishvan2018homeostatic} developed a homeostatic ensemble where the effective temperature emerges from the assumption that the cells maximize entropy while maintaining a homeostatic state. This enabled \cite{shishvan2018homeostatic} to capture the coupling of the biological noise with the environment in an equilibrium setting, i.e. providing the statistics of observations but not the temporal nature of the noise. 

In this paper we extend the active gel approach \tr{\citep{kruse2005generic,JulKruProJoa_pr07}} to include biological noise \tr{as a constitutive material property}. In particular, we explain how this noise can originate from the complex metabolic pathways that drive the ``recharging'' of the ATP hydrolysis chemical reaction providing energy to the cytoskeleton. To do so, we assume that the many internal degrees of freedom that control the rate of the recycling process can be described collectively as an equilibrium system producing Brownian noise. The  Onsager coupling coefficient between mechanics and hydrolysis leading to the presence of active stress then rationalizes the presence of noise in the mechanical stress and thus in the cell shape as a consequence of the fundamental balance laws. Added to this, because of the Onsager symmetry principle, the mechanical state of the cell also feeds back to the dynamics of the energy delivery producing the active stress. The outcome of this procedure is thus an active gel theory wherein  noise not only manifests itself in the active stress but also in the metabolic rate, which becomes inherently coupled to the mechanical environment of the cell.

The paper is constructed as follows. In Section~\ref{sec:theo-framework}, we formulate the mass and momentum conservation laws that characterize the cell mechanical behaviour and build the general thermodynamic framework that we use to derive the constitutive relations of the cell medium. These relations not only involve the connection between stress and deformation in the cell but also describe how the energy is delivered by the cell metabolism to produce an active stress in the system. In Section~\ref{sec:energy_homeo}, we introduce our closure assumption that the complex chemical recycling of the metabolites is described by equilibrium fluctuations whose magnitude is fixed to insure that the chemo-mechanical free energy of a cell is controlled to a fixed target. We then reformulate this general stochastic active gel theory in Section~\ref{sec:sto_crawling_seg} for the simple case of a cell crawling on a straight line. By doing so, we obtain three main results that confirm the applicability of the theory. First we demonstrate the dependence of the metabolic activity on the mechanical environment of the cell. Second  we show that the resulting cell center of mass mean square displacement is in qualitative agreement with experimental results. Third we compute an effective temperature characterizing the long time scale cell fluctuations that is of the correct order of magnitude. Section~\ref{sec:discussion} gathers our conclusions and outlines some potential generalizations of this work.

\section{Close-to-equilibrium thermodynamic framework}\label{sec:theo-framework}

The aim of this section is to derive the constitutive behaviour of the cell modeled as an effective continuum medium. The main challenge is to introduce the regulation by the cell metabolism (i.e. rate of energy delivery) of the active stress actuating the cell skeleton. The open thermodynamic system under consideration \tr{is constituted of the cell in which only the cytoplasm is modeled and a passive semi-infinite visco-elastic (Kelvin-Voigt) substrate on which the cell is adhering. This whole system can exchange work ($W$) and  heat ($Q$)  with an external bath whose temperature and hydrostatic pressure is fixed but the cell, being an open system, can additionally exchange both energy and matter with the bath. We assume that the bath sets the constant temperature $T$ of the cell and the substrate and measure all the mechanical stresses in the system with respect to the bath pressure.}

The cell  occupies the domain $\tr{\omega_t}$ at time $t$ and we denote by $\boldsymbol{x}\in\tr{\omega_t}$  the position of a material point within the cell. The set of points $\boldsymbol{s}\in\partial \tr{\omega_t}$ forms the contour of the cell domain.\tr{This contour can be split into two parts: $\partial\omega_t^b$ representing the contact surface of the cell with the bath and $\partial \omega_t^s$ representing  the contact surface with the substrate. The semi-infinite substrate occupies the domain $\Omega_t$ at time $t$. Its upper boundary  can again be split into two parts with $\partial\Omega_t^c=\partial\omega_t^s$ the contact area with the cell and $\partial\Omega_t^b$ the one with the bath. The rest of the boundary is considered to be static at infinity. The stress-free reference configuration of the substrate (when the cell is absent) is denoted $\Omega_0$. Material points of the substrate are indexed by the spatial variable $\boldsymbol{X}\in\Omega_0$ and can be mapped to the actual configuration by $\boldsymbol{x}=\Psi(\boldsymbol{X},t)\in \Omega_t$.  Based on such mapping, we can define the substrate deformation tensor $\mathbb{F}=\partial_{\boldsymbol{X}} \Psi$ and the Green-Lagrange strain tensor $\mathbb{E}=(\mathbb{F}^T\mathbb{F}-\mathbb{I})/2$, where ${}^T$ denotes the transpose operation and $\mathbb{I}$ is the identity tensor. }   

\paragraph{Conservation laws}

We  write the first principle of thermodynamics as
$$\tr{\dot{U}=\dot{Q}+\dot{W}+\dot{U}_e,}$$
where \tr{the superimposed dot denotes the time derivative}, $U(t)$ is the internal energy of the cell \tr{and the substrate and $U_e(t)$ the  energy that the cell exchanges with the bath.}  The entropy balance reads
$$\dot{S}=\frac{\dot{Q}}{T}+\sigma+\dot{S}_e,$$
where $S(t)$ is the entropy of the \tr{cell and the substrate, $\sigma(t)$ is the entropy production of the full system} and $S_e(t)$ is the entropy which is exchanged between the cell and the bath. The second principle states that the entropy production rate $\sigma\geq 0$.
Combining these two principles  and defining the Helmholtz free energies $F=U-TS$ -free energy of the \tr{cell and its substrate}- and $F_e=U_e-TS _e$ -free energy exchanged between the cell and the bath-, we obtain the dissipation
$$\tr{\mathcal{D}=T\sigma=\dot{W}+\dot{F}_e-\dot{F}\geq 0}.$$
The constitutive equations of the system have to satisfy this fundamental inequality. To compute the dissipation we model the cell cytoskeleton as a \tr{continuum} medium composed of \tr{two phases: a bio-filament meshwork cross-linked by molecular motors that is permeated by  a fluid phase, the cytosol.} The chemical reactions that are necessary to power the motors take place in the cytosol. Next we write the fundamental balance laws governing the behaviour of such a system.

\begin{itemize}
\item \emph{Conservation of momentum.} We denote by $\mathbb{\Sigma}(\boldsymbol{x},t)$ the total Cauchy stress in the cell. This stress splits into a term associated with the filament meshwork $\mathbb{S}(\boldsymbol{x},t)$ and a pressure term associated with the permeating fluid, $P_f(\boldsymbol{x},t)\mathbb{I}$ :
$$\mathbb{\Sigma}=\phi \mathbb{S}-(1-\phi)P_f\mathbb{I}.$$
In the formula above, $\phi(\boldsymbol{x},t)$ denotes the volume fraction of polymer network, which may locally vary. \tr{Note that by \emph{a priori} assuming that the stress in the cytosol is purely hydrostatic, we neglect its bulk viscosity compared to the viscous friction of the fluid on the polymer meswhork filaments as often done in poro-elastic theories \citep{coussy2004poromechanics}. We also introduce $\mathbb{\Sigma}_s$ the Cauchy stress tensor in the substrate.} We assume that all inertial effects can be neglected such that, \tr{denoting $\nabla$ and $\nabla.$ the gradient and divergence operators in the actual configuration,} the force balance laws \tr{within the cell and substrate read
\begin{align*}
-\nabla ((1-\phi)P_f) =-\boldsymbol{f} &\text{ with Boundary Conditions (B.C.) } -(1-\phi)P_f\vert_{\partial \omega_t^b}\boldsymbol{n}=\boldsymbol{t}_e^P\text{ and }-(1-\phi)P_f\vert_{\partial \omega_t^s}\boldsymbol{n}=\boldsymbol{0},\\
\nabla. (\phi\mathbb{S})=\boldsymbol{f} &\text{ with B.C. } \phi \mathbb{S}\vert_{\partial \omega_t^b}\boldsymbol{n}=\boldsymbol{t}_e^S \text{ and } \phi \mathbb{S}\vert_{\partial \omega_t^s}\boldsymbol{n}=\boldsymbol{t}_s,\\
\nabla. \mathbb{\Sigma}_s=\boldsymbol{0} &\text{ with B.C. } \mathbb{\Sigma}_s\vert_{\partial \Omega_t^b}\boldsymbol{n}=\boldsymbol{0} \text{ and } \mathbb{\Sigma}_s\vert_{\partial \omega_t^s}\boldsymbol{n}=\boldsymbol{t}_s,
\end{align*}}
where $\boldsymbol{f}(\boldsymbol{x},t) $ is the interaction force between the fluid and the polymer, \tr{$\boldsymbol{t}_s(\boldsymbol{s},t)$ is the interaction force between the polymer meshwork and the substrate \citep{parsons2010cell}, $\boldsymbol{n}$ the outward unit normal to $\partial\tr{\omega_t}$ and $\boldsymbol{t}_e^P(\boldsymbol{x},t)$ the traction stress externally applied at the boundary of the cell to the fluid and $\boldsymbol{t}_e^S(\boldsymbol{x},t)$ to the meshwork. These tractions may represent for instance those experimentally applied with an optical trap or an AFM \citep{suresh2007biomechanics}.} Therefore the global force balance \tr{in the cell and substrate} read
\tr{\begin{align}
\nabla. \mathbb{\Sigma} =0 &\text{ with B.C. } \mathbb{\Sigma}\vert_{\partial \omega_t^b}\boldsymbol{n}=\boldsymbol{t}_e \text{ and } \mathbb{\Sigma}\vert_{\partial \omega_t^s}\boldsymbol{n}=\boldsymbol{t}_s,\label{e:mec_pb_1}\\
\nabla. \mathbb{\Sigma}_s=0 &\text{ with B.C. } \mathbb{\Sigma}_s\vert_{\partial \Omega_t^b}\boldsymbol{n}=\boldsymbol{0} \text{ and } \mathbb{\Sigma}_s\vert_{\partial \omega_t^s}\boldsymbol{n}=\boldsymbol{t}_s,\label{e:mec_pb_2}
\end{align}
where $\boldsymbol{t}_e=\boldsymbol{t}_e^S+\boldsymbol{t}_e^P$ is the total traction stress externally  applied to the cell. In the absence of body torques, the local balance of torques insures that $\mathbb{\Sigma}$ and $\mathbb{\Sigma}_s$ are symmetric tensors.}
\item \emph{Conservation of mass in the polymer meshwork.} We denote the polymer meshwork  mass density $\rho(\boldsymbol{x},t)$ which obeys the following conservation law \tr{\citep{coussy2004poromechanics}}:
\begin{equation}\label{e:mass_conservation_1}
\partial_t (\phi\rho)+\nabla. (\phi\rho \boldsymbol{v})=R. 
\end{equation}
In the above formula, $\boldsymbol{v}(\boldsymbol{x},t)$ denotes the velocity of the polymer meshwork in the lab frame. The source term $R(\boldsymbol{x},t)$ represents the polymer turnover through its polymerization and depolymerization. We complement \eqref{e:mass_conservation_1} with no flux boundary conditions and consider that the meshwork cannot flow in and out of the cell at our timescale of interest, where the import and export of proteins is negligible and the dry mass of the cell is almost constant. Therefore, introducing $\boldsymbol{v}_b$ the velocity of the domain boundary $\partial\tr{\omega_t}$, we have
\begin{equation}\label{e:bc_polymer}
(\boldsymbol{v}\vert_{\partial \tr{\omega_t}}-\boldsymbol{v}_b).\boldsymbol{n}=0.
\end{equation}
Note that \eqref{e:mass_conservation_1} can be equivalently written as
\begin{equation}\label{e:mass_conservation_2}
\frac{d\phi \rho}{dt}=R-\phi\rho \nabla. \boldsymbol{v},
\end{equation}
where $d./dt=\partial_t.+\boldsymbol{v}.\nabla(.)$ denotes the \tr{material} derivative with respect to the polymer meshwork motion.

\item \emph{Conservation of mass in the cytosol.} Next, we assume that the chemical reaction powering the meshwork is happening inside the cytosol. We will simplify the picture by considering a binary reaction where $a(\boldsymbol{x},t)$ (ATP concentration) is transformed into $b(\boldsymbol{x},t)$ (ADP concentration) inside the cell with a release of energy in the process. The recycling of  $b$ into $a$ is performed by catalytic machines (mitochondria) that use nutrients $n(\boldsymbol{x},t)$ (such as glucose) that enter the cell and are degraded into products $p(\boldsymbol{x},t)$ (such as carbon dioxide) that are expelled out of the cell. Such a recycling happens with a certain stoichiometry whereby one nutrient molecule can recycle $\nu$ molecules of $b$. See Fig.~\ref{fig:scheme_cell} for a schematic of this chemical system.
\begin{figure}[h!]
\centering
\includegraphics[width=0.3\textwidth]{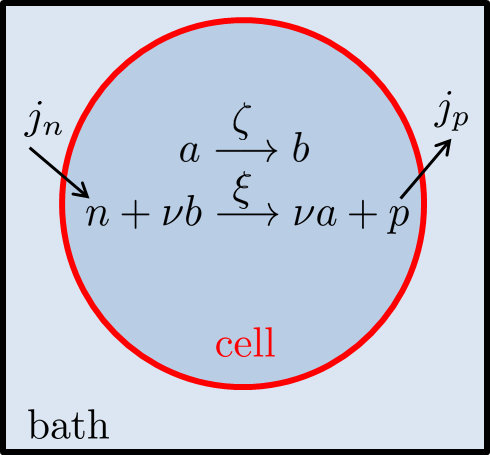}
\caption{\label{fig:scheme_cell} Schematic of the chemical reaction providing the chemical energy necessary to the molecular motors driving the cytoskeleton mechanics.}
\end{figure}
We also consider the monomers associated with the polymer meshwork $g(\boldsymbol{x},t)$ in solution in the cytosol. As a result, we write the mass balance laws in the cytosol as
\begin{align}
\frac{d^f(1-\phi)\rho_f}{dt}=-(1-\phi)\rho_f\nabla.\boldsymbol{v}_f &\text{, ~~} (1-\phi)\rho_f\frac{d^fx_g}{dt}=\nabla.\boldsymbol{J}_g-R, \label{e:mass_balance_cytosol_1}\\
 (1-\phi)\rho_f\frac{d^fx_a}{dt}=\nabla.\boldsymbol{J}_a+(1-\phi)\rho_f\left(\nu\frac{d^f\xi}{dt}-\frac{d^f\zeta}{dt}\right) &\text{, ~~}(1-\phi)\rho_f\frac{d^fx_b}{dt}=\nabla.\boldsymbol{J}_b-(1-\phi)\rho_f\left(\nu\frac{d^f\xi}{dt}-\frac{d^f\zeta}{dt}\right),\label{e:mass_balance_cytosol_2}\\
(1-\phi)\rho_f\frac{d^fx_n}{dt}=\nabla.\boldsymbol{J}_n-(1-\phi)\rho_f\frac{d^f\xi}{dt} &\text{, ~~} (1-\phi)\rho_f\frac{d^fx_p}{dt}=\nabla.\boldsymbol{J}_p+(1-\phi)\rho_f\frac{d^f\xi}{dt},\label{e:mass_balance_cytosol_3}
\end{align}
where $\rho_f$ is the cytosol mass density that we will assume constant (approximately that of water), $\boldsymbol{v}_f(\boldsymbol{x},t)$ is the velocity field of the cytosol, $d^f./dt=\partial_t.+\boldsymbol{v}_f.\nabla(.)$ denotes the total derivative with respect to the fluid motion, $\zeta(\boldsymbol{x},t)$ is the extent of the reaction (\tr{also called progress variable or degree of advancement \citep{de2013non}}) transforming $a$ into $b$, $\xi(\boldsymbol{x},t)$ the extent of the recycling process and $x_{g,a,b,n,p}(\boldsymbol{x},t)=(g,a,b,n,p)/\rho_f$ are the mass fractions of the respective species. These conservation equations are complemented with the boundary conditions
\begin{equation}\label{e:bc_fluid}
(\boldsymbol{v}_f\vert_{\partial \tr{\omega_t}}-\boldsymbol{v}_b).\boldsymbol{n}=0\text{, }\boldsymbol{J}_{a,b}\vert_{\partial \tr{\omega_t}}.\boldsymbol{n}=0 \text{ and } \boldsymbol{J}_{n,p}\vert_{\partial \tr{\omega_t}}.\boldsymbol{n}=j_{n,p}.
\end{equation}
It is therefore assumed for simplicity that water is effectively not flowing in and out the cell at our timescale of interest. Such flows may exist during the fast transient response to  osmotic perturbations of the external environment or during the slow growth of the cell during its cycle \citep{Cadart2019}. The system is open at the level of nutrients and products of the recycling of $b$ into $a$ and we denote the incoming flux of $n$ from the bath by $j_n$ and the outgoing flux of $p$ to the bath  by $j_p$. Note that the total mass of monomers and polymer forming the meshwork is constant:
$$\int_{\tr{\omega_t}}\left(\phi\rho+(1-\phi)g\right)d\boldsymbol{x}=\text{Cst}.$$
\tr{Our model neglects the complexity of the actin turnover dynamics which in reality involves a number of intermediate states as well as other interacting proteins \citep{pollard2016cell}.}
\end{itemize}

\paragraph{Expression for the dissipation}
The rate of \tr{external} work performed on the total system (cell plus substrate) is
\tr{$$\dot{W}=\int_{\partial\omega_t^b}\left(\boldsymbol{t}_e^S.\boldsymbol{v}+\boldsymbol{t}_e^P.\boldsymbol{v}_f\right)d\boldsymbol{s}=\int_{\partial\omega_t^b}\left(\boldsymbol{t}_e.\boldsymbol{v}+\boldsymbol{t}_e^P.\bar{\boldsymbol{v}}_f\right)d\boldsymbol{s},$$
where $\bar{\boldsymbol{v}}_f=\boldsymbol{v}_f-\boldsymbol{v}$ is the relative velocity of the fluid with respect to the polymer meshwork. Using momentum conservation laws \eqref{e:mec_pb_1}-\eqref{e:mec_pb_2}, we obtain that 
$$\int_{\partial\omega_t^b}\boldsymbol{t}_e.\boldsymbol{v}d\boldsymbol{s}=\int_{\omega_t}\mathbb{\Sigma}: \mathbb{D}(\boldsymbol{v})d\boldsymbol{x}-\int_{\partial\omega_t^s}\boldsymbol{t}_s.\boldsymbol{v}d\boldsymbol{s}$$
and
$$0=\int_{\Omega_t}\mathbb{\Sigma}_s: \mathbb{D}(\boldsymbol{v}_s)d\boldsymbol{x}+\int_{\partial\omega_t^s}\boldsymbol{t}_s.\boldsymbol{v}_sd\boldsymbol{s}$$
where $:$ is the canonical scalar product on order two tensors, $\boldsymbol{v}_s(\boldsymbol{x},t)$ is the velocity of the substrate and $\mathbb{D}(.)$ is the symmetric part of the related velocity field. With these two relations, we finally obtain that
$$\dot{W}=\int_{\omega_t}\left(\mathbb{\Sigma}: \mathbb{D}(\boldsymbol{v})-\nabla. [(1-\phi)P_f\bar{\boldsymbol{v}}_f]\right)d\boldsymbol{x}+\int_{\omega_t^s}\boldsymbol{t}_s.\bar{\boldsymbol{v}}_s d\boldsymbol{s}+\int_{\Omega_t}\mathbb{\Sigma}_s: \mathbb{D}(\boldsymbol{v}_s)d\boldsymbol{x}$$
where $\bar{\boldsymbol{v}}_s=\boldsymbol{v}_s-\boldsymbol{v}$.}

To compute the rate of change of the free energy, we postulate that
$$F=\int_{\tr{\omega_t}}\left[ \phi\rho f_{\text{mec}}(\phi\rho,(1-\phi)\rho_f)d\boldsymbol{x}+(1-\phi)\rho_ff_{\text{chem}}(x_g,x_a,x_b,x_n,x_p)d\boldsymbol{x}+\tr{(1-\phi)\rho_f} f_{\text{bio}}(\lbrace\theta_i\rbrace_{i=1..N})\right]d\boldsymbol{x}+\tr{\int_{\Omega_0} f_{\text{sub}}(\mathbb{E})d\boldsymbol{X}},$$
where $f_{\text{mec}}$ is the mechanical free energy per unit mass of polymer, $f_{\text{chem}}$  is the chemical free energy per unit mass of cytosolic solvent, $f_{\text{bio}}$ is associated to the complex biological metabolic pathways controlling the recycling of ADP to ATP \tr{and $f_{\text{sub}}$ is the free energy per unit volume of the visco-elastic substrate (i.e. elastic energy).} On the one hand, we assume that $f_{\text{mec}}$ depends on the internal variables introduced above that control the level of deformation of the cytoskeleton while $f_{\text{chem}}$ depends on the mass fractions of the chemical species in solution in the cytosol. On the other hand, $f_{\text{bio}}$ depends on a large ($N\gg 1$) number of internal degrees of freedom $\lbrace\theta_i(\boldsymbol{x},t)\rbrace_{i=1..N}$  associated with the recycling process of ADP into ATP involving many complex biochemical and regulatory pathways (such as glycolysis and the Krebs cycle). \tr{Finally $f_{\text{mec}}$ only depends on the substrate elastic strain $\mathbb{E}$.}

As a simple example, we may specify some classical form of $f_{\text{mec}}$, $f_{\text{chem}}$ \tr{and $f_{\text{sub}}$} while $f_{\text{bio}}$ is very complex. Typically, in a mechanically linear theory the dependence of $f_{\text{mec}}$ on the polymer and solvent densities is quadratic:
\begin{equation}\label{e:free_ener_mech_exp}
f_{\text{mec}}=\frac{K_{\rho}}{2\bar{\rho}}\left(\frac{\rho\phi-\bar{\rho}}{\bar{\rho}}\right)^2+\frac{K_{\rho_f}}{2\bar{\rho}_f}\left(\frac{\rho_f(1-\phi)-\bar{\rho}_f}{\bar{\rho}_f}\right)^2-K_{\rho}\alpha\frac{\rho\phi\rho_f(1-\phi)}{\bar{\rho}^2\bar{\rho}_f}
\end{equation}
and
\begin{equation}\label{e:free_ener_chem_exp}
f_{\text{chem}}=\sum_{i=g,a,b,n,p}\frac{R_i}{2} \left(\frac{x_i-\bar{x}_i}{\bar{x}_i}\right)^2.
\end{equation}
In the above expressions, $K_{\rho}$ and $K_{\rho_f}$ are compressibility moduli, $\bar{\rho}$ and $\bar{\rho}_f$ equilibrium densities, $\alpha$ a Biot coefficient characterizing the mechanical coupling between the cytoskeletal and cytosol phases and $\bar{x}_{g,a,b,n,p}$ equilibrium mass fractions. The coefficients 
$$R_i=\frac{k_BT\mathcal{N}_A}{m_i}$$
are the specific gas constant of each specie ($k_B$ is the Boltzmann constant, $\mathcal{N}_A$ the Avogadro number and $m_i$ the molar mass of the $\text{i}^{\text{th}}$ specie). \tr{In the same way, a classical quadratic form of the substrate free energy per unit volume is
$$f_{\text{sub}}=\frac{3K_s-2G_s}{6}\text{tr}(\mathbb{E})^2+G_s \mathbb{E}:\mathbb{E}$$
which corresponds to a Saint-Venant Kirchhoff material with bulk modulus $K_s$ and shear modulus $G_s$.} However, these expressions are only examples to fix ideas and the rest of the theory does not rely on this specific choice.

We can  directly compute the rate of change of free energy as 
$$
\dot{F}=\int_{\tr{\omega_t}}\left(Rf_{\text{mec}}+\phi\rho\frac{df_{\text{mec}}}{dt}+(1-\phi)\rho_f\frac{d^ff_{\text{chem}}}{dt}+(1-\phi)\rho_f\frac{d^ff_{\text{bio}}}{dt}\right)d\boldsymbol{x}+\tr{\int_{\Omega_t}\frac{1}{\det \mathbb{F}}\frac{d^sf_{\text{sub}}}{dt}d\boldsymbol{x}},
$$
\tr{where $d^s./dt=\partial_t.+\boldsymbol{v}_s.\nabla(.)$ is the material derivative in the substrate.} We additionally write that the rate of free energy input and output in the system is locally proportional to the rate of delivery and removal of $n$ and $p$  (since the system is closed for all other species),
$$\dot{F}_e=\int_{\partial\tr{\omega_t}}\left(j_n\mu_n^0+j_p\mu_p^0\right)d\boldsymbol{s}=\int_{\tr{\omega_t}}\left(\mu_n^0\nabla.\boldsymbol{J}_n+\mu_p^0\nabla.\boldsymbol{J}_p\right)d\boldsymbol{s}.$$
Here we have introduced the constant free energies per unit mass $\mu_{n,p}^0$ of $n$ and $p$ when they are  brought in ($j_n$ is positive) and removed ($j_p$ is negative) from the system. This is similar to considering the grand canonical ensemble  with a bath that maintains  fixed chemical potentials (in other words, a chemostat).  
Using the chain rule, we obtain
\begin{align*}
&\frac{df_{\text{mec}}}{dt}=\frac{\partial f_{\text{mec}}}{\partial \phi\rho}\frac{d\phi\rho}{dt}+\frac{\partial f_{\text{mec}}}{\partial (1-\phi)\rho_f}\frac{d(1-\phi)\rho_f}{dt} \text{, } \frac{d^ff_{\text{chem}}}{dt}=\sum_{i=g,a,b,n,p}\frac{\partial f_{\text{chem}}}{\partial x_i}\frac{d^fx_i}{dt} \text{, }\\
&\frac{d^ff_{\text{bio}}}{dt}=\underset{i=1..N}\sum\frac{\partial f_{\text{bio}}}{\partial \theta_i}\frac{d^f\theta_i}{dt} \text{ and } \frac{d^sf_{\text{sub}}}{dt}=\frac{\partial f_{\text{sub}}}{\partial \mathbb{E}}:\frac{d^s\mathbb{E}}{dt}.
\end{align*}
Therefore, using the conservation laws \eqref{e:mass_conservation_2}-\eqref{e:mass_balance_cytosol_3}, we can express dissipation as \tr{the sum of products of generalized forces by generalized fluxes}
\begin{align}\label{e:dissipation}
\mathcal{D}&=\int_{\tr{\omega_t}}\left((\mathbb{\Sigma}+\phi P\mathbb{I}+(1-\phi)P_f\mathbb{I} ):\mathbb{D}-(1-\phi)\bar{\boldsymbol{v}}_f\nabla P_f +R(\mu_g-\mu)+\underset{i=g,a,b,n,p}\sum\boldsymbol{J}_i\nabla\mu_i\right.\\
\notag &+(1-\phi)\rho_f\frac{d^f\xi}{dt}\left(\Delta\mu_{np}-\nu\Delta\mu_{ab}\right)+(1-\phi)\rho_f\frac{d^f\zeta}{dt}\Delta\mu_{ab}\\
\notag  &\left.-\underset{i=1..N}\sum\tr{(1-\phi)\rho_f}\frac{d^f\theta_i}{dt}\mu_{\theta_i}\right)d\boldsymbol{x}+\int_{\partial \tr{\omega_t}} \left(j_n(\mu_n^0-\mu_n\vert_{\partial \tr{\omega_t}})+j_p(\mu_p^0-\mu_p\vert_{\partial \tr{\omega_t}})\right)d\boldsymbol{s}\\
&\tr{+\int_{\omega_t^s}\boldsymbol{t}_s.\bar{\boldsymbol{v}}_s d\boldsymbol{s}+\int_{\Omega_t}\left( \mathbb{F}^{-1}\mathbb{\Sigma}_s\mathbb{F}^{-T}-\frac{1}{\det\mathbb{F}}\frac{\partial f_{\text{sub}}}{\partial \mathbb{E}}\right) :\frac{d^s\mathbb{E}}{dt}d\boldsymbol{x}}
\geq 0.\nonumber
\end{align}
In \eqref{e:dissipation}, $P=\phi\rho^2\partial_{\phi\rho}f_{\text{mec}}$ is the thermodynamic pressure in the polymer meshwork, $P_f=\phi\rho \rho_f\partial_{(1-\phi)\rho_f}f_{\text{mec}}$, is the hydrostatic pressure in the permeating fluid,  $\mu=f_{\text{mec}}+P/\rho$ is the Gibbs chemical potential of the polymer, $\mu_{g,a,b,n,p}=\partial_{x_{g,a,b,n,p}}f_{\text{chem}}$ are the chemical potentials of the monomers, ADP, ATP nutrients and products in solution in the cytosol. We define $\Delta\mu_{np}=\mu_n-\mu_p$ and $\Delta\mu_{ab}=\mu_a-\mu_b$. Finally the $\mu_{\theta_i}=\partial_{\theta_i}f_{\text{bio}}$ are the chemical potentials of the biochemical degrees of freedom that fully describe the biochemical processes regulating the recycling of ADP to ATP. 
\tr{The generalized force-flux pairs entering in the dissipation and the conservation laws associated to the generalized fluxes can be summarized in Table~\ref{t:forceflux}.
\begin{table}[h!]
\center
{\renewcommand{\arraystretch}{1.2}
\tr{\begin{tabular}{lcr}
\hline\hline
\emph{generalized force} &\emph{generalized flux}&\emph{conservation law}\\
\hline
$\mathbb{D}$& $\mathbb{\Sigma}+\phi P\mathbb{I}+(1-\phi)P_f\mathbb{I}$&cell momentum balance $\eqref{e:mec_pb_1}$\\
$-\nabla P_f$&$(1-\phi)\bar{\boldsymbol{v}}_f$&fluid conservation $\eqref{e:mass_balance_cytosol_1}_1$\\
$\mu_g-\mu$& $R$&polymer conservation  $\eqref{e:mass_conservation_1}$\\
$\nabla\mu_i$&$\boldsymbol{J}_i$&solutes conservation $\eqref{e:mass_balance_cytosol_1}_2-\eqref{e:mass_balance_cytosol_2}-\eqref{e:mass_balance_cytosol_3}$\\
$\Delta\mu_{np}-\nu\Delta\mu_{ab}$&$(1-\phi)\rho_f d^f\xi/dt$& metabolites conservation  $\eqref{e:mass_balance_cytosol_2}-\eqref{e:mass_balance_cytosol_3}$\\
$\Delta\mu_{ab}$&$(1-\phi)\rho_f d^f\zeta/dt$& metabolites conservation $\eqref{e:mass_balance_cytosol_2}$\\
$-\mu_{\theta_i}$& $\tr{(1-\phi)\rho_f}d^f\theta_i/dt$&recycling processes \\
$\mu_{n,p}^0-\mu_{n,p}$ &$j_{n,p}\vert_{\partial \tr{\omega_t}}$& nutrients and products fluxes $\eqref{e:bc_fluid}_3$\\
$\bar{\boldsymbol{v}}_s$ &$\boldsymbol{t}_s$& cell-substrate traction forces  $\eqref{e:mec_pb_1}-\eqref{e:mec_pb_2}$\\
$d^s\mathbb{E}/dt$&$\mathbb{F}^{-1}\mathbb{\Sigma}_s\mathbb{F}^{-T}-\det\mathbb{F}^{-1}\partial f_{\text{sub}}/\partial \mathbb{E}$& substrate momentum balance  $\eqref{e:mec_pb_2}$\\
\hline\hline
\end{tabular}}}
\caption{\label{t:forceflux}\small \tr{Force-flux pairs entering in the dissipation expression \eqref{e:dissipation}. The fluxes are associated to their specific conservation laws.} }
\end{table}}
\paragraph{Generalized forces-fluxes relations} Close to thermodynamic equilibrium, the generalized fluxes entering in the dissipation can be written as a linear combination (with some symmetries on the kinetic coefficients) of the generalized forces according to Onsager's principle \citep{de2013non}. In particular, we can write the constitutive relations: 
\begin{equation}\label{e:dissipativecase}
\begin{array}{c}
\mathbb{\Sigma}=-\phi P\mathbb{I}-(1-\phi)P_f\mathbb{I} +\eta \mathbb{D}+\chi_{\mathbb{\Sigma} a}\Delta\mu_{ab} \mathbb{I}\text{, } (1-\phi)\bar{\boldsymbol{v}}_f=-\frac{\kappa}{\eta_f}\nabla P_f\text{ and } R=k_{\rho}(\mu_g-\mu)\\
\boldsymbol{J}_{g,a,b,n,p}=M_{g,a,b,n,p}\nabla \mu_{g,a,b,n,p} \text{ and }  j_{n,p}=L_{n,p}(\mu_{n,p}^0-\mu_{n,p}\vert_{\partial \tr{\omega_t}})\\
\rho_f(1-\phi) \frac{d^f\zeta}{dt}=\chi_{a\mathbb{\Sigma}}\nabla.\boldsymbol{v} +k_{a}\Delta\mu_{ab}\\
\rho_f(1-\phi) \frac{d^f\xi}{dt}=k_{n}(\Delta\mu_{np}-\nu\Delta\mu_{ab})-\underset{i=1..N}\sum \lambda_{ni}\mu_{\theta_i}, \\
\forall i=1..N\text{, }  \tr{\rho_f(1-\phi)}\frac{d^f\theta_i}{dt}=-\lambda_{ii}\mu_{\theta_i}+\lambda_{in}(\Delta\mu_{np}-\nu\Delta\mu_{ab}),\\
\tr{\boldsymbol{t}_s=\tilde{\lambda}\bar{\boldsymbol{v}}_s \text{ and } \mathbb{\Sigma}_s =\mathbb{F}\left(\frac{1}{\det\mathbb{F}}\frac{\partial f_{\text{sub}}}{\partial\mathbb{E}}+\eta_s\frac{d^s\mathbb{E}}{dt}\right) \mathbb{F}^{T}.}
\end{array}
\end{equation}
In \eqref{e:dissipativecase}, we have neglected a certain number of cross-couplings  to only retain: 
\begin{enumerate}
\item The active stress driven by the chemical reaction transforming $a$ into $b$ and the corresponding cross term in the equation giving the dynamics of $\zeta$. Because the strain rate is odd under time reversal while $\Delta\mu_{ab}$ is even, we have $\chi_{\mathbb{\Sigma} a}=-\chi_{a\mathbb{\Sigma}}$ \citep{kruse2005generic}. The active stress is isotropic  here but it can be given deviatoric components  by involving in the free energy a polarity field as in more general liquid crystal theories \citep{kruse2005generic}. We also neglect for sake of simplicity the fact that the growth properties of the polymer meshwork can be actively controlled by the ATP hydrolysis which could be easily accounted for by introducing a cross-coupling term between $R$ and $\Delta\mu_{ab}$ (and its symmetric counterpart).
\item  The regulating action of the micro-variables $\theta_i$ on the kinetics of the recycling process of ADP to ATP. Therefore the $\theta_i$ may be understood as all the molecular degrees of freedom involved in the running of mitochondria. As the sign reversal signature is the same for all coupled variables in this case, we have the symmetry relations  and $\lambda_{ni}=\lambda_{in}$. 
\end{enumerate}
In \eqref{e:dissipativecase}, \tr{we have also neglected any type of material anisotropy and}  $\eta$ denotes the viscosity of the polymer meshwork, $\eta_f$($\ll \eta$ in practice) is the viscosity of the permeating fluid, $\kappa$ is the permeability (in [$\text{m}^2$]) of the polymer meshwork, $k_{\rho}$ is the rate of renewal of the network,  $M_{g,a,b,n,p}$ are the Fickian mobilities of the various species in solution, $L_{n,p}$ are the permeabilities of the cell membrane to the nutrients and the products, $k_a$ is the rate of the chemical reaction transforming  ATP into ADP and $k_n$ the rate of turnover of nutrients into products. \tr{Finally, $\tilde{\lambda}$ is a friction coefficient of the cell with its substrate and $\eta_s$ is the substrate viscosity. For a purely elastic substrate, we recover the general relation giving the Cauchy stress as a function of the elastic energy $\mathbb{\Sigma}_s =\frac{1}{\det\mathbb{F}}\mathbb{F}\frac{\partial f_{\text{sub}}}{\partial\mathbb{E}} \mathbb{F}^{T}$ while for small deformations of substrate, we obtain the usual linear relation $\mathbb{\Sigma}_s =\frac{\partial f_{\text{sub}}}{\partial\mathbb{E}}+\eta_s\frac{d^s\mathbb{E}}{dt}$.} 

\section{Energy homeostasis}\label{sec:energy_homeo}

Combining the last two equations of \eqref{e:dissipativecase} to eliminate $\mu_{\theta_i}$, we obtain the kinetics of the metabolic recycling of ADP into ATP:
$$\rho_f(1-\phi) \frac{d^f\xi}{dt}=\left( k_{n}-\bar{\lambda}_n\right) \left( \Delta\mu_{np}-\nu \Delta\mu_{ab}\right)+\underset{i=1..N}\sum\frac{\lambda_{ni}}{\lambda_{ii}} \tr{\rho_f(1-\phi)}\frac{d^f\theta_i}{dt}
\text{ where, }\bar{\lambda}_n=\underset{i=1..N}\sum\frac{\lambda_{ni}^2}{\lambda_{ii}}.$$
At this stage, we make the strong modeling assumption that, because the $\mu_{\theta_i}({\lbrace\theta_i\rbrace}_{i=1..N})$ are complex coupled functions, the  large  system of equations ruling the dynamics of the internal degrees of freedom $\theta_i$ is \tr{also complex and, to effectively describe the collective $\theta_i$ dynamic, we suppose that these variables are independent and identically distributed stochastic processes.} In such a case, according to the central limit theorem, 
$$X(\boldsymbol{x},t)=\tr{\rho_f(1-\phi)}\underset{i=1..N}\sum\frac{\lambda_{ni}}{\lambda_{ii}}\theta_i(\boldsymbol{x},t)$$ 
converges to a biased Brownian motion characterized by a certain mean $X_0$ and variance $\Theta$. The mean is irrelevant here as the $\theta_i$ only enter the problem under a time derivative. This constitutive assumption that represents the recycling of metabolites as an equilibrium system is of course  questionable for many systems as, for instance, chemical reactions are known to produce coloured rather than white noise \citep{sekimoto2010stochastic}.  But this simple closure provides a noisy chemical recycling dynamics 
$$\rho_f(1-\phi) \frac{d^f\xi}{dt}=\bar{k}_n\left( \Delta\mu_{np}-\nu\Delta\mu_{ab}\right) +\frac{d^fX}{dt},$$
where 
$$ \bar{k}_n=k_{n}-\bar{\lambda}_n.$$
Thus, the stochasticity of the energy delivery  entails a stochastic behaviour of $x_a$ and $x_b$ resulting in a stochastic mechanical system through the active stress. This in particular implies that cell motility, which is driven by its cytoskeleton, is controlled by the recycling of ATP as we shall demonstrate in the following section. Correlation between fluctuations of cell shape changes and ATP concentration have been experimentally demonstrated by \cite{suzuki2015spatiotemporal}.

Although we have reduced the $\theta_i$ biochemical variables involved in the recycling of ADP to ATP to a single effective parameter $\Theta$, we have not specified yet how these variables collectively regulate the energy recycling process. To do so, we first define  the chemo-mechanical cell free-energy  as  
$$F_{cm}=F_{\text{mec}}+F_{\text{chem}}=\int_{\tr{\omega_t}}\phi\rho f_{\text{mec}}(\phi\rho,(1-\phi)\rho_f) d\boldsymbol{x}+\int_{\tr{\omega_t}} (1-\phi)\rho_ff_{\text{chem}}(x_g,x_a,x_b,x_n,x_p)d\boldsymbol{x}.$$
The rate of change of $F_{\text{mec}}$ is associated with the energetic cost for the cell to perform its mechanical tasks like changing its shape or moving its center of mass \citep{recho2014optimality}. Following the idea presented in \citep{shishvan2018homeostatic,buskermolen2019entropic} we postulate that changes of $F_{\text{mec}}$ are compensated by changes of $F_{\text{chem}}$ such that the total chemo-mechanical free energy $F_{cm}$ is fixed independently of the external mechanical loading. This is a constitutive assumption that we associate with the biological idea of  cell energy homeostasis.  In other words, the cell machinery functions to maintain at a constant level the energy resource that can be employed by the cytoskeleton.  Thus, denoting by $\mathbb{E}(.)$ the ensemble averaging over the noise fluctuations, the amplitude of the fluctuations $\Theta$ is set by the non local (in both space and time) constraint:
\begin{equation}\label{e:energy_homeosta}
\underset{\tr{\tilde{t}}\rightarrow \infty}{\lim}\frac{1}{\tr{\tilde{t}}}\int_0^{\tr{\tilde{t}}}\mathbb{E} (F_{cm})\,dt=  F_0, 
\end{equation}
where $F_0$ is a constant average value over both time and stochastic fluctuations of the chemo-mechanical free energy,  which is evaluated from that of a suspended cell (i.e. in the absence of geometrical confinement or external force). 

The full model is therefore composed of the mass and momentum conservation laws in which the general constitutive relations of the active medium are given by the Forces-fluxes Onsager relations \eqref{e:dissipativecase}. These relations are simplified by assuming that the large number of internal metabolic degrees of freedom that characterize the recycling of the energy delivery process to the cytoskeleton can be treated as an equilibrium reservoir that  controls  the global free energy of the cell to a target value by the homeostatic constraint \eqref{e:energy_homeosta}.  Because the active stress entering in the Onsager relations stems from a dissipative coupling between cytoskeleton mechanics and ATP hydrolysis through $\chi_{\mathbb{\Sigma}a}$, the active stress becomes a stochastic variable that is controlled by the metabolism. Reciprocally, the rate of recycling of the energy delivery to the cytoskeleton is influenced by cell mechanics in general, and by the mechanical environment of the cell in particular.    

\section{Example: an active stochastic segment}\label{sec:sto_crawling_seg}

Rather than giving a general discussion of the theoretical framework developed above, we illustrate and explain its implications in the simple case of a one-dimensional \tr{infinitely thin} segment of active gel crawling along a straight \tr{rigid} track. Our aim is to spell out the general coupling discussed above in this special case and show how it can help to understand the stochastic nature of cell motility. In this situation $\tr{\omega_t}=[l_-(t),l_+(t)]$ where $l_-<l_+$ denote the positions of the cell fronts. To simplify the formulation of the problem, we introduce the traveling coordinate $y=x-l_-\in [0,L]$ where $L=l_+-l_-$ is the gel length. In this new coordinate system $\partial_x=\partial_y$ and $\partial_t\vert_{y}=\partial_t\vert_{x}+\dot{l}_-\partial_x\vert_{x}$, where the superimposed dot denotes the time derivative of the front position. 
\begin{figure}[h!]
\centering
\includegraphics[width=0.35\textwidth]{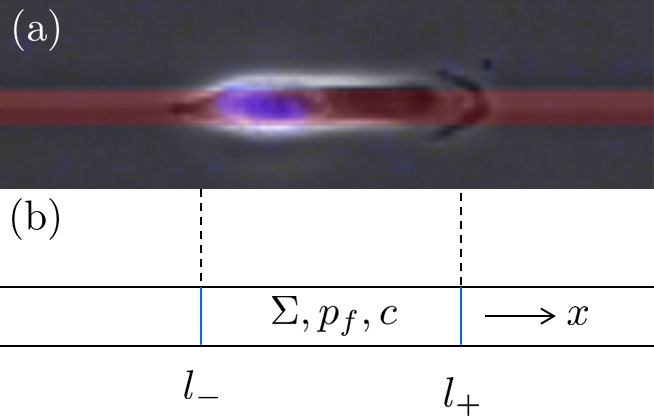}
\caption{\label{fig:} (a) Human renal adenocarcinoma migrating on a fibronectin coated track from \cite{maiuri2012first}. (b) Scheme of the crawling active gel segment.}
\end{figure}

\subsection{Mechanics of the active gel segment}
\tr{As the track is considered to be infinitely stiff, the substrate velocity vanishes ($\boldsymbol{v}_s=0$) and we obtain from the last line of \eqref{e:dissipativecase}  that $\boldsymbol{t}_s=-\tilde{\lambda} \boldsymbol{v}$. In the context of a thin film approximation (the cell height $h\ll L $), this friction results in a body force $-\lambda v$ where the rescaled friction coefficient is $\lambda=\tilde{\lambda}/h$. See \cite{roux2016prediction} for details.}
From \eqref{e:mec_pb_1}, the global force balance therefore reads
\begin{equation}\label{e:f_bal}
\partial_y\Sigma=\lambda v.
\end{equation}
 As we do not consider any additional external force, \eqref{e:f_bal} is associated with the boundary conditions:
\begin{equation}\label{e:f_bal_bc}
\Sigma\vert_{0,L}=0.
\end{equation}
The first relation of \eqref{e:dissipativecase} provides the constitutive behaviour of the gel 
\begin{equation}\label{e:const_be}
\Sigma=-\phi P-(1-\phi)P_f +\eta \partial_yv+\chi_{\Sigma a}\Delta \mu_{ab}.
\end{equation}
To specify the pressure terms $P$ and $P_f$, we consider the free energy expression \eqref{e:free_ener_mech_exp} and obtain
$$P=K_{\rho}\phi \frac{\rho^2}{\bar{\rho}^2}\left[\frac{\rho\phi-\bar{\rho}}{\bar{\rho}}-\alpha\frac{\rho_f(1-\phi)}{\bar{\rho}_f}\right] \text{ and } P_f=\phi \frac{\rho\rho_f}{\bar{\rho}_f^2}\left[K_{\rho_f}\frac{\rho_f(1-\phi)-\bar{\rho}_f}{\bar{\rho}_f}-\alpha K_{\rho}\frac{\bar{\rho}_f\rho\phi}{\bar{\rho}^2}\right].$$
Next, we make the realistic assumptions that the polymer meshwork is highly compressible while the volume fraction is fixed and hence consider the limit where $K_{\rho}=0$, $K_{\rho_f}=\infty$ and $\rho_f(1-\phi)=\bar{\rho}_f$ while the product $K_{\rho_f}(\rho_f(1-\phi)-\bar{\rho}_f)$ remains finite. These assumptions entail  that the volume fraction is fixed $\phi=1-\bar{\rho}_f/\rho_f$ and $P_f$ is a Lagrange multiplier determined by the solvent mass conservation relation $\text{(\ref{e:mass_balance_cytosol_1})}_1$ which reduces to:
$$\partial_yv_f=0.$$
Hence, using the boundary conditions \eqref{e:bc_fluid} the fluid velocity is a constant equal to the common velocity of the moving fronts $v_f=V=\dot{l}_-=\dot{l}_+$ where $V(t)$ denotes the common front velocity.

Combining this simple relation with the Darcy law obtained from \eqref{e:dissipativecase}, we derive the hydrostatic pressure field:
\begin{equation}\label{e:Darcy_moving}
(1-\phi)(V-v)=-\frac{\kappa}{\eta_f}\partial_y P_f.
\end{equation}
From the boundary conditions associated with the absence of flow of polymer outside the cell membrane \eqref{e:bc_polymer}, we have
$$v\vert_{0,L}=V \text{ so that }\partial_yP_f\vert_{0,L}=0.$$

Another useful consequence of the infinite compressibility of the polymer meshwork and fixed volume fraction is that the mechanical free energy reduces to zero
\begin{equation}\label{e:free_ener_mech_simpl}
f_{\text{mec}}=0.
\end{equation}
Note that as the constitutive behaviour \eqref{e:const_be} no longer involves $P$, the mass balance equation for the polymer $\text{(\ref{e:mass_conservation_1})}_1$ uncouples from the mechanical problem and the meshwork density $\rho$ can be reconstructed afterwards (in tandem with the concentration of monomers $g$) when the velocity field $v$ is determined (see for instance \citep{recho2013contraction, recho2015mechanics}). For simplicity, we suppose that $\rho$ is fixed at a constant corresponding to the local chemical equilibrium between the polymerization of the meshwork and its depolymerization. More specifically, plugging the assumption of constant $\rho$ into  $\text{(\ref{e:mass_conservation_1})}_1$ with the turnover rate $R$ given by \eqref{e:dissipativecase}, we obtain
$$k_{\rho}^{-1}\phi\rho\partial_yv=\mu_g-\mu $$
which, if the rate of turnover $k_{\rho}$ is much larger than the rate of transport, reduces to $\mu(\rho)=\mu_g(\bar{x}_g)$. This last relation fixes the value of $\rho$ at a constant if the diffusion of $g$ is large enough such that its concentration is homogeneous in the whole cell: $x_g=\bar{x}_g$.

With these simplifying but realistic assumptions, the mechanical problem describing the cytoskeleton can be formulated in the compact form
\begin{equation}\label{e:mec_problem_crawling}
\left\lbrace \begin{array}{c}
-\frac{\eta}{\lambda}\partial_{yy}\Sigma+\Sigma=-(1-\phi)P_f+\chi_{\Sigma a}\Delta \mu_{ab}\\
\partial_{yy}\left[\frac{\lambda\kappa}{\eta_f}P_f-(1-\phi)\Sigma \right]=0.
\end{array}\right. 
\end{equation}
 The boundary conditions associated to \eqref{e:mec_problem_crawling} are
\begin{equation}\label{e:mec_problem_crawling_bc}
\Sigma\vert_{0,L}=0\text{ and } \partial_yP_f\vert_{0,L}=0.
\end{equation}
Once \eqref{e:mec_problem_crawling}-\eqref{e:mec_problem_crawling_bc} is solved, the cell front dynamics can be computed from the integration of\eqref{e:Darcy_moving} over the whole segment:
\begin{equation}\label{e:front_dyna_cell}
(1-\phi)V(t)L=-\frac{\kappa}{\eta_f}\left[P_f(L,t)-P_f(0,t) \right].
\end{equation}
When $\Delta\mu_{ab}$ is fixed, \eqref{e:mec_problem_crawling}-\eqref{e:mec_problem_crawling_bc} reduces to an isotropic active gel model  similar to that studied in \cite{JulKruProJoa_pr07}. However, the ingredient of the permeation of the cytosol in the cytoskeleton \citep{alt1999cytoplasm,callan2013active, kimpton2015poroviscoelastic} has been added to this classical model.  The model equations \eqref{e:mec_problem_crawling} describe a non-polarized cell. Several mechanisms leading to a spontaneous cell mechanical polarization have been discussed in the literature (See for instance \cite{callan2013active,recho2013contraction, blanch2013spontaneous, tjhung2012spontaneous,edelstein2013simple, giomi2014spontaneous}) and could be added to this framework. We have deliberately left this important effect aside to focus on the coupling of mechanics with metabolism and the ensuing stochastic nature of the cell motion. To do so, system \eqref{e:mec_problem_crawling} is now coupled with a paradigmatic model of the cell metabolism that dynamically sets $\Delta\mu_{ab}$. This coupling introduces stochasticity in the deterministic mechanical model as detailed below. 

\subsection{Dynamics of the energy delivery}

From the Onsager relations \eqref{e:dissipativecase} and the mass balance equations \eqref{e:mass_balance_cytosol_2}-\eqref{e:mass_balance_cytosol_3} of the solute chemical species $a,b,n$ and $p$, we obtain
\begin{equation}\label{e:metabol}
\begin{array}{c}
\rho_f(1-\phi)\partial_tx_a=M_a\partial_{yy}\mu_a+\left[\nu\bar{k}_n\left( \Delta\mu_{np}-\nu \Delta\mu_{ab}\right)+\nu\partial_tX +\chi_{\Sigma a}\partial_yv-k_a\Delta\mu_{ab}  \right] \\
\rho_f(1-\phi)\partial_tx_b=M_b\partial_{yy}\mu_b-\left[\nu\bar{k}_n\left( \Delta\mu_{np}-\nu \Delta\mu_{ab}\right)+\nu\partial_tX +\chi_{\Sigma a}\partial_yv-k_a\Delta\mu_{ab}  \right]\\
\rho_f(1-\phi)\partial_tx_n=M_n\partial_{yy}\mu_n-\left[\bar{k}_n\left( \Delta\mu_{np}-\nu \Delta\mu_{ab}\right)+\partial_tX \right]  \\
\rho_f(1-\phi)\partial_tx_p=M_p\partial_{yy}\mu_p+\left[\bar{k}_n\left( \Delta\mu_{np}-\nu \Delta\mu_{ab}\right)+\partial_tX \right].  
\end{array}
\end{equation} 
Using some technical assumptions that rely on the fast diffusion of these chemical species and that are explained in detail in \ref{sec:appendix0}, we can simplify \eqref{e:metabol} to the single partial differential equation
 \begin{equation}\label{e:dyna_delta_mu}
\begin{array}{c}
\partial_t\Delta\mu_{ab}=D_{ab}\partial_{yy}\Delta\mu_{ab}+\frac{2\kappa\chi_{\Sigma a}}{\alpha_{ab} (1-\phi)\eta_f}\partial_{yy}P_f-\frac{2k_{na}}{\alpha_{ab}}\Delta\mu_{ab}+\frac{2\nu\bar{k}_n}{\alpha_{ab}}\Delta\mu_{np}^0+\frac{2\nu}{\alpha_{ab}} \Gamma,
\end{array}
\end{equation}
which encapsulates the kinetics of the metabolism that provides energy to the molecular motors actuating the active gel. In \eqref{e:dyna_delta_mu}, $D_{ab}$ (estimated in Table~\ref{t:valpar}) is an effective diffusion coefficient of the metabolites in the cytosol and $\Gamma(y,t)=\partial_tX(y,t)$
is a  Gaussian process satisfying
$$\mathbb{E}( \Gamma(y,t))=0 \text{ and } \mathbb{E}( \Gamma(y,t)\Gamma(y',t'))=2\Theta\text{min}(y,y')\delta(t-t').$$
Note that the physical dimension of $\Theta$ is thus given by $[\Theta]=\text{kg}^2\text{m}^{-7}\text{s}^{-1}$.

Mechanics and metabolism are coupled by the Onsager coefficient $\chi_{\Sigma a}$ which enters in both \eqref{e:mec_problem_crawling} and \eqref{e:dyna_delta_mu}. In the absence of metabolic noise in the system, $\Gamma=0$ and the solution of the problem is trivial:
\begin{equation}
\Delta \mu_{ab}=\Delta \mu_{ab}^0=\nu \frac{\bar{k}_n}{k_{na}}\Delta \mu_{np}^0.
\end{equation}
This leads to an homogeneous distribution of stresses within the active gel segment:
$$\Sigma=\Sigma^0=0 \text{ and } P_f=P_f^0=\chi_{\Sigma a}\Delta \mu_{ab}^0/(1-\phi).$$ 
In the presence of metabolic noise, we write $\Delta \mu_{ab}=\Delta \mu_{ab}^0+\delta \mu_{ab}$, $\Sigma=\Sigma^0+\delta \Sigma$ and $P_f=P_f^0+\delta P_f$ where, given the linearity of equation \eqref{e:dyna_delta_mu},  $\delta \mu_{ab}$ is a stochastic variable with a zero mean ($\mathbb{E}(\delta \mu_{ab})=0$) whose amplitude is set by the noise amplitude. Therefore, to close the problem, it remains to apply the energetic constraint that sets the value of $\Theta$. In the framework of the active segment, the energy homeostasis assumption \eqref{e:energy_homeosta} reads:
\begin{equation}\label{eq:homeo_constraint_segment_simpl}
 \underset{\tr{\tilde{t}}\rightarrow \infty}{\lim}\frac{1}{L\tr{\tilde{t}}}\int_0^{\tr{\tilde{t}}}\int_0^L  \mathbb{E}(\delta \mu_{ab}^2)dydt =\mu_0^2,
\end{equation}
where  $\mu_0^2$ is a constant explicitly related to $F_0$ (See \ref{sec:appendix0}).

\subsection{Solution of the chemo-mechanical problem}

Combining the mechanical problem \eqref{e:mec_problem_crawling} and \eqref{e:mec_problem_crawling_bc} and the kinetics of the metabolism controlling the active stress \eqref{e:dyna_delta_mu}, we obtain the coupled chemo-mechanical stochastic problem:
\begin{equation}\label{eq:final_pb}
\left\lbrace \begin{array}{c}
-\frac{\eta}{\lambda}\partial_{yy}\delta\Sigma+\delta\Sigma=-(1-\phi)\delta P_f+\chi_{\Sigma a}\delta \mu_{ab}\\
\partial_{yy}\left[\frac{\lambda\kappa}{\eta_f}\delta P_f-(1-\phi)\delta \Sigma \right]=0\\
\partial_t\delta\mu_{ab}=D_{ab}\partial_{yy}\delta\mu_{ab}+\frac{2\kappa\chi_{\Sigma a}}{\alpha_{ab} (1-\phi)\eta_f}\partial_{yy}\delta P_f-\frac{2k_{na}}{\alpha_{ab}}\delta\mu_{ab}+\frac{2\nu}{\alpha_{ab}} \Gamma
\end{array}\right. 
\end{equation}
with boundary conditions
\begin{equation}\label{eq:final_pb_bc}
\delta\Sigma\vert_{0,L}=0\text{, } \partial_y\delta P_f\vert_{0,L}=0  \text{ and } \partial_y\delta\mu_{ab}\vert_{0,L}=0.
\end{equation}
The parameters entering in \eqref{eq:final_pb}-\eqref{eq:final_pb_bc} can be estimated based on various experiments, see Table \ref{t:valpar}. 

\begin{table}
\center
\begin{tabular}{lcr}
\hline\hline
name & symbol & typical value \\ 
\hline
cytoskeleton viscosity & $\eta$ & $10^3$ Pa s \citep{JulKruProJoa_pr07,Rubinstein2009}\\
cytosol viscosity & $\eta_f$ & $2\times 10^{-3}$ Pa s \citep{moeendarbary2013cytoplasm}\\
cytoskeleton permeability & $\kappa$ & $2\times 10^{-16}$ $\text{m}^2$ \citep{moeendarbary2013cytoplasm}\\
solid volume fraction & $\phi$ & $0.25$  \citep{moeendarbary2013cytoplasm}\\
energy production per unit mass & $ \Delta\mu_{ab}^0$ &$ 10^5$ $\text{J}.\text{kg}^{-1}$ \citep{JulKruProJoa_pr07}\\
contractility & $\chi_{\Sigma a} \Delta\mu_{ab}^0$ &$10^3$ Pa \citep{JulKruProJoa_pr07,Rubinstein2009}\\
viscous friction coefficient & $\lambda$ &$10^{15}$ Pa~s~$\text{m}^{-2}$ \citep{JulKruProJoa_pr07, barnhart2011adhesion} \\
cell length & $L$ &$ 10^{-5}$ m\\
energy conversion coefficent & $\alpha_{ab}$ & $10^{-5} \text{ kg}^2\text{m}^{-3}\text{J}^{-1}$ [see \ref{sec:appendix0}]\\
diffusion of ATP/ADP & $D_{ab}$ &$ 10^{-12}$ $\text{m}^{2}\text{s}^{-1}$ [see \ref{sec:appendix0}]\\
number of ATP recycled with one nutrient & $\nu$ & 30 \citep{Alberts2002}\\
rate of the ATP to ADP reaction in motors & $k_a/\alpha_{ab}$ & $25\text{s}^{-1}$  \citep{howard2001mechanics}\\
rate of the ATP recycling & $\bar{k}_n/\alpha_{np}$ & $0.01\text{s}^{-1}$ \citep{skog1982energy}\\
effective rate (see \eqref{eq:effective_rate}) & $k_{na}/\alpha_{ab}$ & $25\text{s}^{-1}$ \\
\hline\hline
\end{tabular}
\caption{\label{t:valpar}\small Estimates of material and kinetic coefficients entering in the chemo-mechanical model. It is however important to keep in mind that some of these biophysical parameters (such as cytoskeleton viscosity or the viscous friction coefficient \cite{barnhart2011adhesion} for instance) can vary over several orders of magnitudes depending on the cell type. }
\end{table}

To solve the above linear but non-local problem, we first consider the first two equations in \eqref{eq:final_pb} with their associated boundary conditions to obtain the following expression 
$$\partial_{yy}P_f[ \delta\mu (y,t)]=\frac{\lambda  \chi_{\Sigma a} \left(\frac{\eta  \tr{\Lambda} ^2}{\lambda }-1\right) }{\eta  (1-\phi )}\left(\frac{\tr{\Lambda}  \left(\int_y^L  \psi (u,y) \delta \mu (u,t)\, du+\int_0^y \psi (y,u)\delta \mu (u,t)  \, du\right)}{L \tr{\Lambda}  \left(\frac{\eta
    \tr{\Lambda} ^2}{\lambda }-1\right) \sinh (L \tr{\Lambda} )+4 \sinh ^2\left(\frac{L \tr{\Lambda} }{2}\right)}-\delta \mu (y,t)\right),$$
where the interaction kernel reads
$$\psi (y,u)=L \tr{\Lambda}  \left(\frac{\eta  \tr{\Lambda} ^2}{\lambda }-1\right) \cosh (u \tr{\Lambda} ) \cosh (\tr{\Lambda}  (L-y))+2 \sinh \left(\frac{L \tr{\Lambda} }{2}\right) \cosh \left(\tr{\Lambda}  \left(\frac{L}{2}+u-y\right)\right).$$
\tr{This kernel is obtained using the standard method of the variation of constants for a second order problem.}
In the  expressions above, we have introduced the  hydrodynamic wavelength  
$$\tr{\Lambda}=\sqrt{\frac{\kappa  \lambda +\eta_f(1-\phi )^2}{\eta  \kappa }},$$
a generalization of the quantity introduced in \cite{JulKruProJoa_pr07}. Next, we write
$$\delta \mu(y,t)=\sum_{k=0}^{\infty}\delta\mu_k(t)w_k(y) \text{ with }\delta\mu_k(t)=\int_0^L\delta\mu(y,t)w_k(y)dy$$
and
$$\delta \Gamma(y,t)=\sum_{k=0}^{\infty}\Gamma_k(t)w_k(y) \text{ with }\Gamma_k(t)=\int_0^L\Gamma(y,t)w_k(y)dy$$
to project the last equation of \eqref{eq:final_pb} on the Hilbert-Schmidt basis 
$$w_k(y)=\sqrt{\frac{2}{L}}\cos\left( \frac{k\pi}{L}y\right).$$
After an exponentially decaying transient, the steady state modes of $\delta\mu$ are given by the relations (See \ref{sec:appendix1}.)
\begin{equation}\label{e:deltamu_modes}
\forall\,k\geq 0,\left\lbrace \,\begin{array}{c}
\delta\mu_{2k}(t)=\frac{2}{\alpha_{ab}}\int_0^t\text{e}^{d_{2k}(t-u)}\Gamma_{2k}(u)du \\
\delta\mu_{2k+1}(t)=\frac{2}{\alpha_{ab}}\int_0^t\text{e}^{d_{2k+1}(t-u)}\Gamma_{2k+1}(u)du+\frac{2}{\alpha_{ab}s}\int_0^t\text{e}^{d_{2k+1}(t-u)}\left( \text{e}^{s(t-u)}-1\right)  \beta_{2k+1}\sum_{l=0}^{\infty}\gamma_{2l+1}\Gamma_{2l+1}(u)du,
\end{array}\right.
\end{equation}
where the three sequences entering in \eqref{e:deltamu_modes} are given by:
$$d_k=-\left( \frac{k^2\pi^2}{\tau_d}+\frac{2}{\tau_r}+\frac{2 \pi ^2 k^2 }{\tau_a  \left(\pi ^2 k^2+ f^2\right)}\right)\text{, } \gamma_k=-\frac{4 \sqrt{2/L}  f^2  \text{sech}\left(f/2\right) }{\tau_a \left(\pi ^2 k^2+f^2 \right) \left(p f- f +2
     \tanh(f/2)\right)}\text{ and }\beta_k=\frac{2 \sqrt{2L}  f\cosh(f/2)}{ \pi ^2 k^2+f^2}$$
and the scalar $s$ reads:
$$ s=\frac{(f-\sinh (f)) \text{sech}^2\left(f/2\right)}{\tau_a \left(f p-f+2 \tanh \left(f/2\right)\right)}.$$
In the above expression, we have introduced the characteristic timescales representing the diffusion time, the reaction time and a transport time mediated by the molecular motors
$$\tau_d= \frac{L^2}{D_{ab}}\text{, }\tau_r= \frac{\alpha_{ab}}{k_{na}}\text{ and }\tau_a= \frac{\eta\alpha_{ab}}{\chi_{\Sigma a}^2}$$
and two non-dimensional parameters :
$$f=\tr{\Lambda} L=\frac{L}{l_0}\left(Q_{\lambda}+\frac{1}{Q_{\eta}} \right)^{1/2} \text{ and } p=\frac{\eta\tr{\Lambda}^2}{\lambda}=1+\frac{1}{Q_{\lambda}Q_{\eta}}.$$ 
For a more transparent physical interpretation, we have also expressed $f$ and $p$ as a function of $l_0=\sqrt{\kappa}/(1-\phi)$ the characteristic permeation length scale, $Q_{\eta}=\eta/\eta_f$ the non-dimensional ratio of the cytoskeleton to cytosol viscosity and $Q_{\lambda}=\lambda \kappa/(\eta (1-\phi)^2)$ which is another non-dimensional parameter representing the external friction due to the environment divided by the internal friction of the cytoskeleton in the cytosol. The cell environment properties are therefore all encapsulated in this last parameter $Q_{\lambda}$. \tr{Our simple model of non-polarizable cell motility on a one dimensional track  essentially depends on a few non-dimensional parameters only, namely $Q_{\lambda}$, $Q_{\eta}$, two independent ratios of the three characteristic timescales $\tau_d$, $\tau_r$ and $\tau_a$ and the ratio of the two lengthscales $L/l_0$. } 

Finally, we use the relations \eqref{e:deltamu_modes} to set $\Theta$ according to the global energy homeostasis constraint \eqref{eq:homeo_constraint_segment_simpl}.
Using the noise statistics property
\begin{equation}\label{e:noise_autocorr}
\mathbb{E}(\Gamma_k(t)\Gamma_l(t'))=2\Theta\int_0^L\int_0^L \text{min}(y,y')\delta(t-t')w_k(y)w_k(y')dydy'=2\Theta \delta(t-t')\delta(l-k)\left\lbrace\begin{array}{c}
\frac{2L^2}{3}\text{ if } k=0\\
\frac{L^2}{k^2\pi^2}\text{ if } k\geq 1
\end{array} \right.,
\end{equation}
we obtain the relation setting $\Theta$ as a function of the constant value $\mu_0^2$,
\begin{align}\label{e:homeo_constraint_2}
\mu_0^2=&\underset{\tr{\tilde{t}}\rightarrow \infty}{\lim}\frac{1}{L\tr{\tilde{t}}}\int_0^{\tr{\tilde{t}}} \sum_{k=0}^{\infty}\mathbb{E}(\delta\mu_k^2)dt=\frac{\Theta L \tau_r}{\alpha_{ab}^2}\left( \frac{4}{3}-\sum_{k=1}^{\infty}\frac{1}{\tau_r(2\pi k)^2d_{2k}}-\right. \\
&\left. 4\sum_{k=0}^{\infty}\frac{s^2-2(s+d_{2k+1})\beta_{2k+1}\gamma_{2k+1}+3sd_{2k+1}+2d_{2k+1}^2+\bar{s}\beta_{2k+1}^2}{\tau_r(\pi(2k+1))^2d_{2k+1}(s+d_{2k+1})(s+2d_{2k+1})}\right),\nonumber
\end{align}
where
$$\bar{s}=\sum_{k=0}^{\infty}\gamma_{2k+1}^2.$$

\begin{figure}[h!]
\centering
\includegraphics[width=0.5\textwidth]{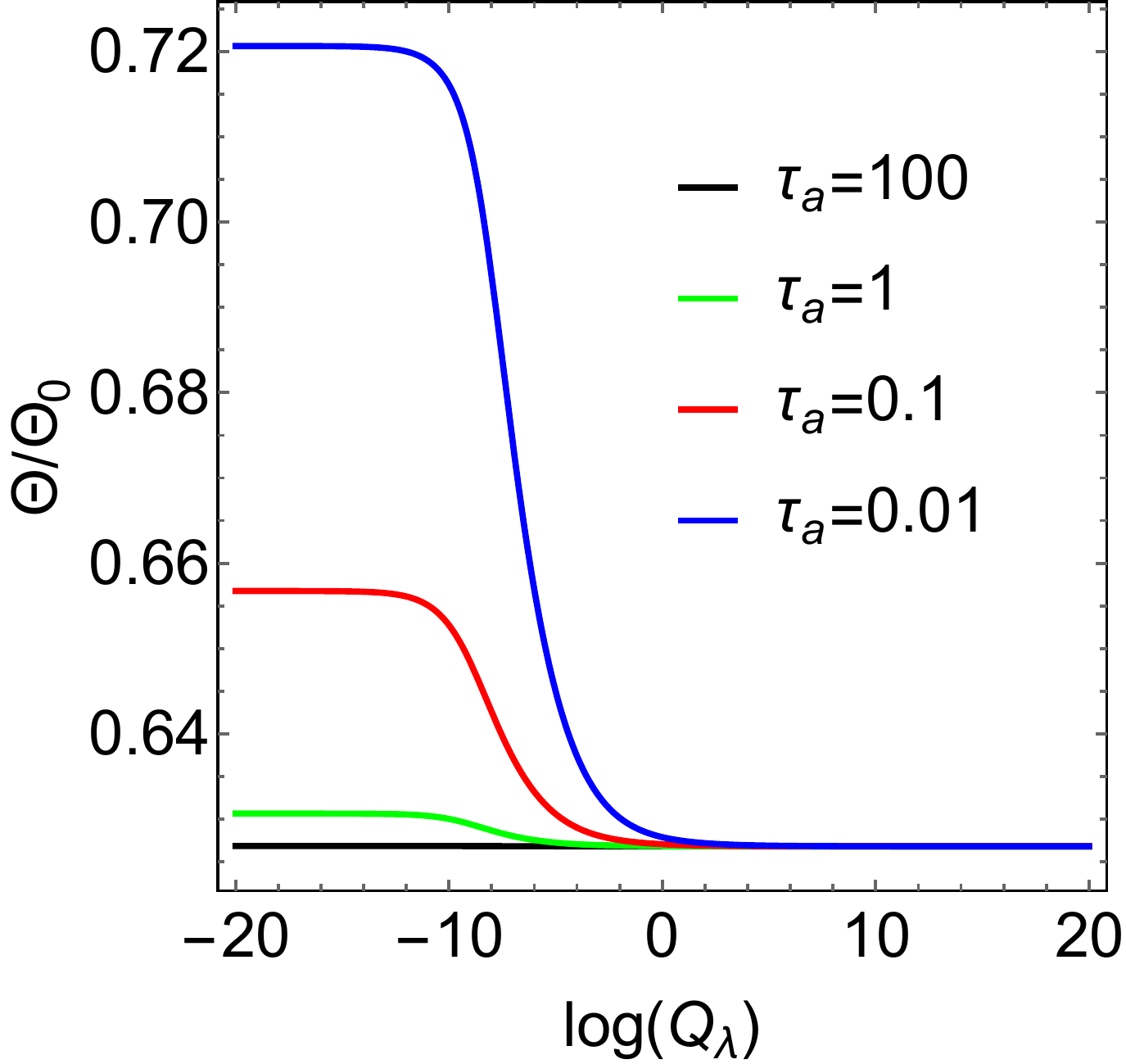}
\caption{\label{fig:Theta_Q_lambda}  Amplitude of the fluctuations $\Theta$ controlling the energy homeostasis  as a function of $Q_{\lambda}$, the friction with the external environment. Using the parameters in Table \ref{t:valpar}, we set: $\tau_d\simeq 100$s, $\tau_r\simeq 0.04$s,  $L/l_0\simeq 530$ and $Q_{\eta}\simeq 10^6$.}
\end{figure}

As a consequence, depending on the mechanical properties of the environment, the biochemical regulation of the recharging of ADP into ATP  is influenced to comply with the homeostatic energy constraint. This dependence of the stochastic fluctuations on the environment is rooted in the cell mechanical activity. Indeed if $\chi_{\Sigma a}=0$ (i.e. $\tau_a=\infty$), the relation fixing $\Theta$ is independent of $\lambda$ as in these conditions,  
$$\mu_0^2=\frac{\Theta L \tau_r}{\alpha_{ab}^2}\left( \frac{1}{48} \left(\frac{6 \tau_r}{\tau_d}-3 \sqrt{2} \sqrt{\frac{\tau_r}{\tau_d}} \left(4 \tanh \left(\sqrt{\frac{\tau_d}{2\tau_r}}\right)+\coth \left(\sqrt{\frac{\tau_d}{2\tau_r}}\right)\right)+77\right) \right) \underset{\tau_r/\tau_d\rightarrow 0}\rightarrow \frac{77\Theta L \tau_r}{48\alpha_{ab}^2}.$$  
We show on Fig.~\ref{fig:Theta_Q_lambda} the value of $\Theta$ normalized by $\Theta_0=\alpha_{ab}^2\mu_0^2/(L\tau_r)$ for several choices of $\tau_a$. When $\tau_a$ is large compared to $\tau_r$ (i.e. the cell contractility is small), $\Theta$ is almost constant as a function of the external viscous friction $Q_{\lambda}$ and assumes the value computed above. Outside of this limit, the value of $\Theta$ as a function of $Q_{\lambda}$ emerging from the energy homeostatic constraint displays two plateau regions connected  by a region of  decrease. This region where $\Theta$ varies corresponds to a physiological range $Q_{\lambda}\in [10^{-10},1]$.    The fact that, depending on the cell type, the glycolysis pathway is affected by the rigidity of the environment  was recently demonstrated experimentally by \cite{park2020mechanical}. 
\FloatBarrier
As experimental access to  $\Theta$ is difficult, we now formulate more easily testable predictions by computing the center of mass fluctuations of the cell in our motility model.

\subsection{Stochastic cell motion}

Based on formula \eqref{e:front_dyna_cell}, we obtain the following expression relating cell velocity and $\Delta \mu_{ab}$:
\begin{equation}\label{e:velocity_to_Deltamu}
V(t)=\frac{\chi_{\Sigma a}  \tr{\Lambda} ^2 \text{csch}\left(\frac{ \tr{\Lambda} L }{2}\right) \int_0^L  \sinh \left(\tr{\Lambda} (\frac{L}{2}- u)\right) \Delta \mu_{ab} (u,t) \, du}{\eta   \tr{\Lambda} ^3 L-\lambda   \tr{\Lambda} L +2 \lambda  \tanh(\tr{\Lambda} L/2)}.
\end{equation}
Clearly, when $\Delta\mu_{ab}=\Delta\mu_{ab}^0$ is constant, the cell velocity vanishes ($V=0$) because our model does not sustain spontaneous polarization as in \cite{recho2013contraction}. This is due to the fact that, for simplicity and to stay in the strict Onsager framework where kinetic coefficients are constants, we have not considered that the active contractile stress depends on the local concentration of molecular motors. However, in the presence of the metabolic fluctuations $\Gamma$, the cell velocity  still undergoes non-trivial fluctuations as the same expression \eqref{e:velocity_to_Deltamu} relates $V$ with $\delta\mu_{ab}$. Our model is designed to investigate the properties of such fluctuations.

\begin{figure}[h!]
\centering
\includegraphics[width=0.8\textwidth]{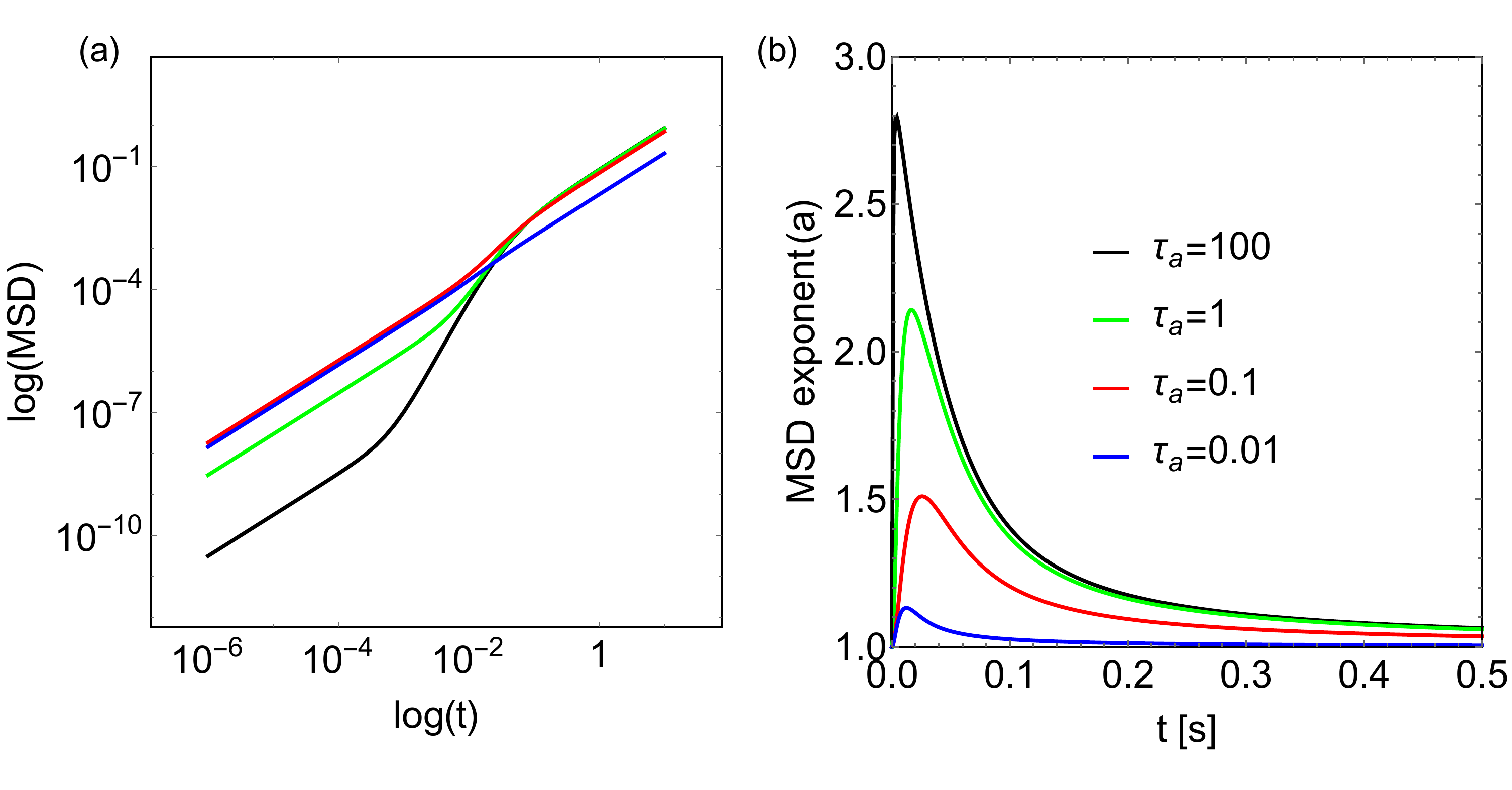}
\caption{\label{fig:MSD} (a) Scaling in a log-log plot of the MSD with time for several values of $\tau_a$. Two diffusive regimes at short and long timescales are connected by a super-diffusive regime at intermediate timescales. (b) Evolution with time of the exponent of the MSD $a=t\partial_t\text{MSD}(t)/\text{MSD}(t)$. Using the parameters in Table \ref{t:valpar}, we set: $\tau_d\simeq 100$s, $\tau_r\simeq 0.04$s,  $L/l_0\simeq 530$,  $Q_{\eta}\simeq 10^6$ and $Q_{\lambda}\simeq 10^{-4}$.}
\end{figure}

To quantify the persistence and randomness of the cell motion \citep{stokes1991migration,dieterich2008anomalous,petrie2009random,maiuri2015actin}, we follow the classical path and define the variation of the cell position from its initial location by
$$\delta X(t)=\int_0^tV(u)\, du,
\text{ and its associated mean square displacement }
\text{MSD}(t)=\mathbb{E}(\delta X(t)^2).$$
Plugging \eqref{e:deltamu_modes} into \eqref{e:velocity_to_Deltamu} we express the velocity as a function of the $\delta \mu$ modes:
$$V(t)=\frac{2 \sqrt{2/L}\chi_{\Sigma a}\tr{\Lambda} }{\lambda}\sum_{k=0}^{\infty}\vartheta_{2k+1}\delta \mu_{2k+1}(t)\text{ where }\vartheta_{k}=\frac{ f^2  }{  \left(f^2+\pi ^2 k^2\right) \left(f p-f+2 \tanh \left(f/2\right)\right)}.$$
Then, using the solution for the odd modes in \eqref{e:deltamu_modes} and the noise statistical property \eqref{e:noise_autocorr}, we can express the MSD as  
$$\text{MSD}(t)=\frac{64\Theta L \chi_{\Sigma a}^2\tr{\Lambda}^2}{\lambda^2\alpha_{ab}^2\pi^2} \int_0^t\int_0^t\int_0^t\int_0^t\delta (v-v')\text{H}(u-v)\text{H}(u'-v')\text{tr}(\mathbb{A}_O(u-v,u'-v'))dudu'dvdv',$$
where H denotes the Heaviside function and the operator $\mathbb{A}_O$ reads
$$\mathbb{A}_O(u-v,u'-v')=\left(\mathbb{I}+\frac{\text{e}^{s(u-v)}-1}{s}\mathbb{M}_O^T\right) \text{e}^{\mathbb{D}_O(u-v)}\overline{\vartheta}_O\overline{\vartheta}_O^T\text{e}^{\mathbb{D}_O(u'-v')}\left(\mathbb{I}+\frac{\text{e}^{s(u'-v')}-1}{s}\mathbb{M}_O\right)\bar{\mathbb{D}}_O^{-1},$$
with the various quantities entering in the above expression defined in \ref{sec:appendix1}. We show in Fig.~\ref{fig:MSD} the typical scaling exponent $a$ of the MSD as a function of time: $\text{MSD}(t) \sim t^{a}$. The cell motion is diffusive ($a=1$) at both very short time and long timescales and undergoes a transition to super-diffusive ($1 \leq a\leq 2$) and hyper-ballistic ($a\geq 2$) motion in between. \tr{In \cite{dieterich2008anomalous}, based on experimental measurements of the positions of cells moving on a two-dimensional substrate, the authors report a similar qualitative behavior where the exponent of the MSD $a$ reaches a maximum at an intermediate timescale.}

The effective diffusion coefficient for cell motion  at a long time scale defined as 
$$D_{\text{eff}}^{\text{cell}}=\lim_{t\rightarrow \infty}\frac{\text{MSD}(t)}{2t},$$
takes a simple form:
\begin{equation}\label{eq:diff_coeff_long_time}
D_{\text{eff}}^{\text{cell}}=\frac{4\alpha_{ab}l_0^2\mu_0^2}{\tau_r\tau_a\eta}\frac{\Theta}{\Theta_0}\frac{p}{Q_{\lambda}} \sum_{k=0}^{\infty}\left( \frac{\vartheta_{2k+1}}{\pi(2k+1)d_{2k+1}}\right)^2.
\end{equation}
Based on such a diffusion coefficient, we can compute a cell effective temperature using the Einstein-Smoluchowski relation:
\begin{equation}\label{e:ES_relation}
T_{\text{eff}}^{\text{cell}}=\frac{\xi_{\text{eff}}^{\text{cell}}D_{\text{eff}}^{\text{cell}}}{ k_B},
\end{equation}
where $\xi_{\text{eff}}^{\text{cell}}$ is an effective cell friction coefficient with its environment that can be related to the local one by $\xi_{\text{eff}}^{\text{cell}}=\lambda w L h$ where $w$ is the cell track width (so $w L$ is the cell contact area) and $h$ is the cell thickness. This effective temperature corresponds to the absolute temperature needed to obtain the observed fluctuations of the cell center of mass if the cell were a passive solid object in a thermal bath \tr{for which the fluctuation-dissipation relation \eqref{e:ES_relation} would apply. See \cite{selmeczi2008cell} for more details on this phenomenological link.}
We thus obtain
\begin{equation}\label{eq:temp_long_time}
T_{\text{eff}}^{\text{cell}}=\frac{4wLh\alpha_{ab}\mu_0^2}{k_B\tau_r\tau_a}\frac{\Theta}{\Theta_0}p \sum_{k=0}^{\infty}\left( \frac{\vartheta_{2k+1}}{\pi(2k+1)d_{2k+1}}\right)^2.
\end{equation}
\begin{figure}[h!]
\centering
\includegraphics[width=0.8\textwidth]{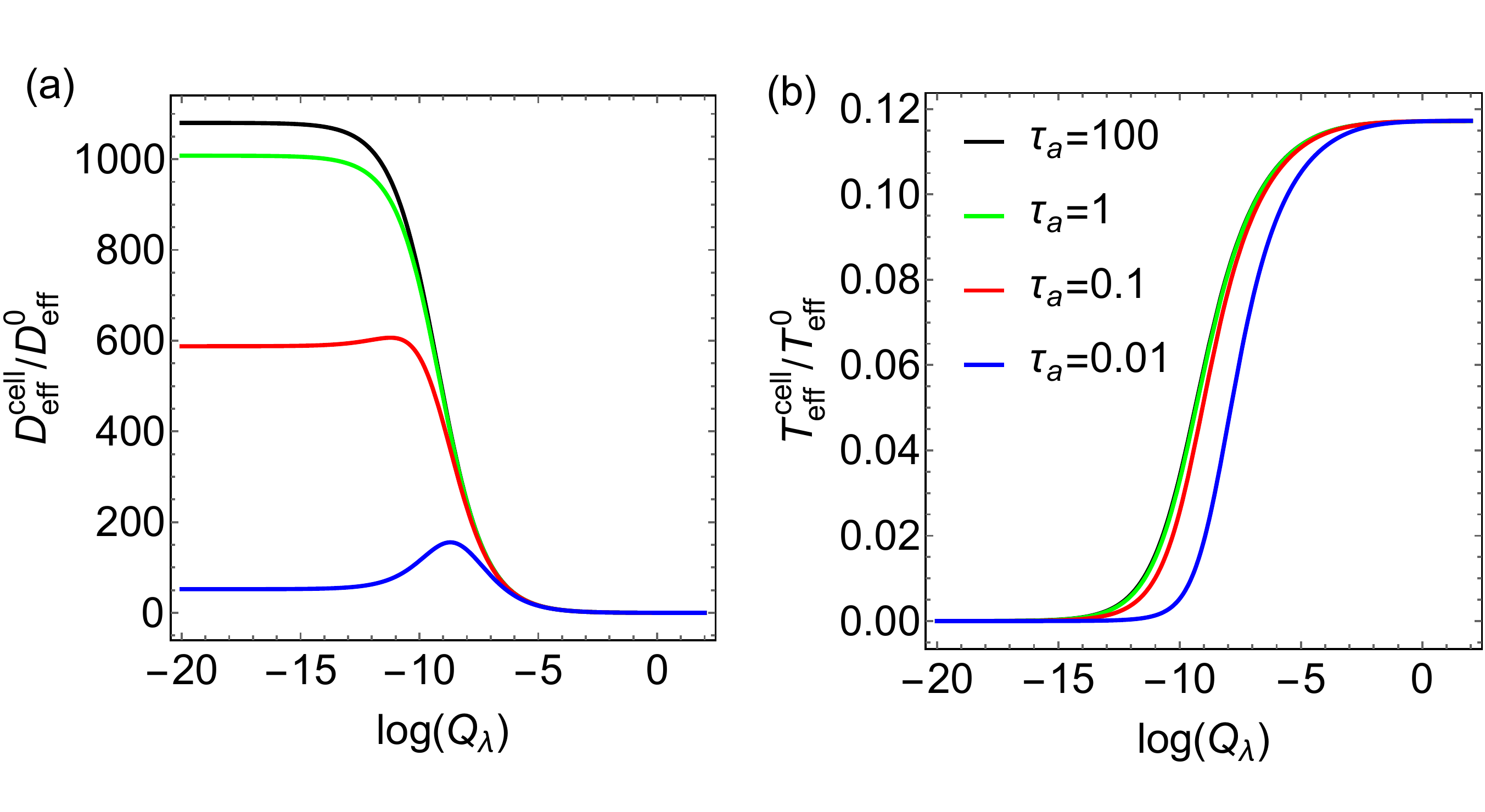}
\caption{\label{fig:diffu_temp} (a) Effective diffusion coefficient of the cell along its track as a function of the external friction. The normalization constant is $D_{\text{eff}}^{0}=4\alpha_{ab}\mu_0^2\tau_r^2l_0^2/(\tau_a\eta)$.   (b) Effective temperature of the cell along its track as a function of the external friction. The normalization constant is $T_{\text{eff}}^{0}=4wLh\alpha_{ab}\mu_0^2\tau_r^2/(k_B\tau_a)$. Using the parameters in Table \ref{t:valpar}, we set: $\tau_d\simeq 100$s, $\tau_r\simeq 0.04$s,  $L/l_0\simeq 530$ and  $Q_{\eta}\simeq 10^6$. }
\end{figure}
Based on the rough estimates reported in Table \ref{t:valpar}, we can estimate using formula \eqref{eq:diff_coeff_long_time} the value of $\mu_0$ such that the cell effective diffusive motion is realistic:  $D_{\text{eff}}^{\text{cell}}\simeq 10\mu \text{m}^2s^{-1}$ \citep{estabridis2018cell,prahl2020predicting,maiuri2015actin}. Doing so, we obtain $\mu_0\simeq 2\times 10^6$~J/kg. Multiplying this quantity by the mass of ATP contained in a single cell ($0.5\times 10^{-15}$~kg for an ATP concentration of $1$mM), we estimate the total free energy of the cell to be $ 10^{-9}$ J which is line with the value reported in \cite{shishvan2018homeostatic}. We can also estimate using \eqref{eq:temp_long_time} the typical effective temperature of a cell. Using the estimates of Table~\ref{t:valpar} and $\mu_0\simeq 2\times 10^6$ J/kg, we obtain $T_{\text{eff}}^{\text{cell}}\simeq 10^{10}$ K, again in agreement with \cite{shishvan2018homeostatic}. Our model therefore provides a realistic explanation for the cell fluctuations of the center of mass which are rooted in the stochastic nature of the energy delivery to the molecular motors actuating the cytoskeleton.

We show on Fig.~\ref{fig:diffu_temp} the dependence of $D_{\text{eff}}^{\text{cell}}$ and $T_{\text{eff}}^{\text{cell}}$  on the  parameter $Q_{\lambda}$ quantifying the friction with the environment.  When $Q_{\lambda}$ is very large, the effective diffusion of the cell goes to zero while it reaches a plateau independent of the value of $Q_{\lambda}$ for low friction. These two limiting behaviors are connected by a region (containing the range of physiological values of $Q_{\lambda}$)  where $D_{\text{eff}}^{\text{cell}}$ changes with $Q_{\lambda}$. These variations are not necessarily monotonic depending of the value of $\tau_a$. The effective temperature of the cell increases with $Q_{\lambda}$ to reach a plateau region at high friction, where temperature becomes independent of the mechanical environment.

\section{Discussion}\label{sec:discussion}

We have presented a linear close-to-equilibrium formalism of cell mechanics coupling the active behaviour of its cytoskeleton with the metabolic pathways recycling the molecules delivering chemical energy into the system. A key assumption of our model is that this complex recycling process is effectively described by an equilibrium reservoir producing fluctuations whose magnitude is fixed by \emph{energy homeostasis}, i.e. to insure that the chemo-mechanical free energy of the cell remains fixed regardless of the external mechanical conditions. As a result, we obtain a stochastic active gel model of the cytoskeleton which generalizes the deterministic approach of \cite{JulKruProJoa_pr07}.

We then apply this formalism to the model problem of a cell randomly migrating along a straight one-dimensional track. This simple example reveals that the fluctuations of the cell position and the fluctuations of the recycling process of the chemo-mechanical energy delivery are mechanically coupled through an Onsager cross coefficient at the origin of active contractile stress in the system. In particular, the model can explain how the metabolic fluctuations can be influenced by the mechanical forces applied on the cell in order to fulfill the energy homeostasis constraint. It also predicts, in agreement with experiments, that the statistical fluctuations of the cell position are diffusive at  short and long timescales (with different effective diffusion coefficients) and that these two regimes are connected by a super-diffusive regime at intermediate timescales. The magnitude of the long timescale cell diffusion coefficient is in agreement with a rough estimate of the total cell energy resources. We also show how such a diffusion is modified by the external mechanical properties of the environment. In our simple model these are encapsulated in a single friction coefficient only. Depending on the magnitude of the active gel contractility, this diffusion coefficient can have a non-trivial maximum corresponding to an optimal friction coefficient where the cell mobility is maximal.
 
The linearity of our cell migration model allows us to obtain some results analytically, without having to rely on numerics but does not allow us to investigate some interesting instabilities. In particular, if we were to augment the model by considering distributions of molecular motors as in  \cite{recho2015mechanics}, it would be able to generate a spontaneous polarization of the cell. The integration of such a property in the present formalism can be achieved by supposing that $\chi_{\Sigma a}(c)$ depends on $c(y,t)$ the motor local concentration. It would lead to interesting  predictions linking not only the velocity standard deviation but also the velocity average value and the statistics of the repolarization events with the stochasticity of the energy renewal processs and the energy homeostasis. \tr{This could serve as a basis to justify from an active gel standpoint the effective phase-field theories presented in \cite{prentice2016directional, alonso2018modeling, moreno2020modeling} and would also provide a potential theoretical background to justify the biological noise introduced in these theories.} In the same vein, it would be natural to assume that the kinetic coefficient controlling the cell metabolism  $\bar{k}_n(\Delta \mu)$ depends on $\Delta \mu$ as, for instance, energy is required for the performance of certain steps in the glycolisis or Krebs cycle. This would have the consequence that  the environment mechanics will not only control the standard deviation of the rate at which glycolisis runs but also the mean of the rate as observed experimentally by \cite{park2020mechanical}. More work will be needed in the future to explore these issues.

\section{Acknowledgments}
This research project was initiated during the Summer School on ``Cell Mechanobiology'' held in September 2018 at  CISM (Udine).
A.D.S. acknowledges support from ERC (Advanced Grant 340685-MicroMotility). R.M.M. acknowledges support by the MRSEC Program NSF under Award No. DMR 1720256 (IRG-3). P.R. acknowledges support from a CNRS MOMENTUM grant. The authors are grateful to L. Truskinovsky for insightful and critical comments.

\appendix 

\section{Justification of expression \eqref{e:dyna_delta_mu}. }\label{sec:appendix0}

Given the quadratic form of the chemical free energy \eqref{e:free_ener_chem_exp} we can directly relate the mass fractions with their associated chemical potentials such that $\partial_tx_i= \bar{x}_i\partial_t\mu_i/R_i$
and we can rewrite \eqref{e:metabol} using only the chemical potentials as variables:
\begin{equation}\label{e:metabol_mu}
\begin{array}{c}
\alpha_a\partial_t\mu_a=M_a\partial_{yy}\mu_a+\chi_{\Sigma a} \partial_yv-k_{na}\Delta\mu_{ab}+\nu\bar{k}_n\Delta\mu_{np}+\nu\partial_tX\\
\alpha_b\partial_t\mu_b=M_b\partial_{yy}\mu_b-\chi_{\Sigma a} \partial_yv+k_{na}\Delta\mu_{ab}-\nu\bar{k}_n\Delta\mu_{np}-\nu\partial_tX\\
\alpha_n\partial_t\mu_n=M_n\partial_{yy}\mu_n+\nu\bar{k}_n\Delta\mu_{ab}-\bar{k}_n\Delta\mu_{np}-\partial_tX \\
\alpha_p\partial_t\mu_p=M_p\partial_{yy}\mu_p-\nu\bar{k}_n\Delta\mu_{ab}+\bar{k}_n\Delta\mu_{np}+\partial_tX,  
\end{array}
\end{equation}
where we have introduced the notations 
\begin{equation}\label{eq:effective_rate}
k_{na}=\nu^2\bar{k}_n+k_a \text{ and }\alpha_{i}=\rho_f(1-\phi)\frac{\bar{x}_{i}}{R_{i}}.
\end{equation}
The boundary conditions associated to \eqref{e:metabol_mu} (see \eqref{e:bc_fluid}) are no flux for $a$ and $b$: $\partial_y\mu_{a,b}\vert_{0,L}=0$ and imposed chemical potentials for $n$ and $p$: $\mu_{n,p}\vert_{0,L}=\mu_{n,p}^0$. These last Dirichlet conditions  correspond in \eqref{e:dissipativecase} to the physical limit where the nutrients and products permeation coefficients $L_{n,p}$ are large  while the products $j_{n,p}=L_{n,p}(\mu_{n,p}\vert_{0,L}-\mu_{n,p}^0)$, representing the incoming and outcoming fluxes for $n$ and $p$, remain finite. 

We first simplify the system \eqref{e:metabol_mu} by approximating $\alpha_{a}\simeq \alpha_{b}\simeq \alpha_{ab}=(\alpha_{a}^{-1}+\alpha_{b}^{-1})^{-1}$ and  $\alpha_{n}\simeq \alpha_{p}\simeq \alpha_{np}=(\alpha_{n}^{-1}+\alpha_{p}^{-1})^{-1}$. These assumptions rely on the value of the molar masses ($m_{ATP}=507$g/mol, $m_{ADP}=427$g/mol, $m_{glucose}=180$g/mol and $m_{CO_2}=44$ g/mol) and an estimate of the various species concentrations in the cell ($[ATP]=0.7$mM, $[ADP]=0.2$mM, $[glucose]=0.5$mM and $[CO_2]=1$mM for Dictyostelium discoideum \citep{albe1990cellular,blombach2015co2})  and lead to $ \alpha_{ab}\simeq 10^{-5} \text{ kg}^2\text{m}^{-3}\text{J}^{-1}$ and $ \alpha_{np}\simeq 5\times 10^{-7} \text{ kg}^2\text{m}^{-3}\text{J}^{-1}$.

Similarly, we suppose that $M_a \simeq M_b \simeq M_{ab}=(M_{a}^{-1}+M_{b}^{-1})^{-1}$ and $M_n \simeq M_p \simeq M_{np}=(M_{n}^{-1}+M_{p}^{-1})^{-1}$ since the Stokes radii of $a$ and $b$ ($r_{ATP}\simeq r_{ADP}=7\times 10^{-10}$m) and $n$ and $p$ ($r_{glucose}=3.8\times 10^{-10}$m and $r_{CO_2}=2\times 10^{-10}$m) are about the same.  The resulting  diffusion coefficients are $D_{np}=M_{np}/\alpha_{np}=k_{\text{B}}T/(6\pi r_{np}\eta_f)\simeq 1.5\times 10^{-9}\text{m}^2.\text{s}^{-1}$ and $D_{ab}=M_{ab}/\alpha_{ab}=k_{\text{B}}T/(6\pi r_{ab}\eta_f)\simeq 3\times 10^{-10}\text{m}^2.\text{s}^{-1}$. However, ATP and ADP are highly reactive which results in a trapping that effectively reduces their diffusion coefficient by one or two orders of magnitude \citep{saks2003heterogeneity}. 
Having assumed that $n$ and $p$ as well as $a$ and $b$ have the same effective mobility in the cell, we conclude that 
$$S_{ab}=\frac{\mu_a+\mu_b}{2}\text{ and }S_{np}=\frac{\mu_n+\mu_p}{2}$$
are constants fixed by the initial conditions and we can reduce \eqref{e:metabol_mu} to a system of only two stochastic partial differential equations involving the difference of chemical potentials
\begin{equation}\label{e:metabol_mu_reduce}
\begin{array}{c}
\frac{\alpha_{ab}}{2}\partial_t\Delta\mu_{ab}=\frac{M_{ab}}{2}\partial_{yy}\Delta\mu_{ab}+\chi_{\Sigma a}\partial_yv-k_{na}\Delta\mu_{ab}+\nu\bar{k}_n\Delta\mu_{np}+\nu \Gamma\\
\frac{\alpha_{np}}{2}\partial_t\Delta\mu_{np}=\frac{M_{np}}{2}\partial_{yy}\Delta\mu_{np}+\nu\bar{k}_n\Delta\mu_{ab}-\bar{k}_n\Delta\mu_{np}-\Gamma,
\end{array}
\end{equation}
where $\Gamma(y,t)=\partial_tX(y,t)$
is a  Gaussian process satisfying
$$\mathbb{E}( \Gamma(y,t))=0 \text{ and } \mathbb{E}( \Gamma(y,t)\Gamma(y',t'))=2\Theta\text{min}(y,y')\delta(t-t').$$

We now further simplify the above system \eqref{e:metabol_mu_reduce} by neglecting the reaction term in $\text{(\ref{e:metabol_mu_reduce})}_2$ compared to diffusion. This reduction is based on the fact that the diffusive length scale $(M_{np}/\bar{k}_n)^{1/2}\simeq 400\mu$m  is an order of magnitude larger than the typical cell length ($10\mu$m). To estimate $\bar{k}_n$ we consider the global rate of ATP turnover in the cell $k_n/\alpha_{np}$ to be $0.01 \text{ s}^{-1}$ \citep{skog1982energy}. We therefore approximate $\Delta\mu_{np}=\Delta\mu_{np}^0+\bar{\Gamma}$, where $\bar{\Gamma}$ is a Gaussian noise satisfying the diffusion equation $\partial_t\bar{\Gamma}=D_{np}\partial_{yy}\bar{\Gamma}-(2/\alpha_{np})\Gamma$
with Dirichlet boundary conditions. Since this equation is linear, it is clear that $\bar{\Gamma}$ remains Gaussian with a zero mean while its covariance can be directly related to $\Theta$. Plugging this expression for $\Delta\mu_{np}$ into $\text{(\ref{e:metabol_mu_reduce})}_1$, we  obtain
 \begin{equation}
\begin{array}{c}
\frac{\alpha_{ab}}{2}\partial_t\Delta\mu_{ab}=\frac{M_{ab}}{2}\partial_{yy}\Delta\mu_{ab}+\chi_{\Sigma a} \partial_yv-k_{na}\Delta\mu_{ab}+\nu\bar{k}_n\Delta\mu_{np}^0+\nu (\Gamma+\bar{k}_n\bar{\Gamma})
\end{array}
\end{equation}
which, differentiating relation \eqref{e:Darcy_moving} and using again the fact that  $M_{np}/(\bar{k}_nL^2)\gg 1$ can be put in the final form:
 \begin{equation}
\begin{array}{c}
\partial_t\Delta\mu_{ab}=D_{ab}\partial_{yy}\Delta\mu_{ab}+\frac{2\kappa\chi_{\Sigma a}}{\alpha_{ab} (1-\phi)\eta_f}\partial_{yy}P_f-\frac{2k_{na}}{\alpha_{ab}}\Delta\mu_{ab}+\frac{2\nu\bar{k}_n}{\alpha_{ab}}\Delta\mu_{np}^0+\frac{2\nu}{\alpha_{ab}} \Gamma.
\end{array}
\end{equation}

In the framework of the active segment, the energy homeostasis assumption \eqref{e:energy_homeosta} reads:
\begin{equation}\label{eq:homeo_constraint_segment}
(1-\phi)\rho_f S\sum_{i=g,a,b,n,p}\frac{R_i}{2}\underset{T\rightarrow \infty}{\lim}\frac{1}{T}\int_0^T\int_0^L  \mathbb{E}\left(\left( \frac{x_i-\bar{x}_i}{\bar{x}_i}\right)^2\right)dydt =F_0,
\end{equation}
where $S$ is a unit surface equating the height of the cell (typically a few microns) multiplied by the thickness of the track (typically again a few microns). Based on the fact that $x_g=\bar{x}_g$, $S_{ab}$ and $S_{np}$ are fixed and $\mathbb{E}(\delta \mu_{ab})=0$, we can rewrite the condition \eqref{eq:homeo_constraint_segment} in a form that only involves $\mathbb{E}(\delta \mu_{ab}^2)$:
\begin{equation}
 \underset{T\rightarrow \infty}{\lim}\frac{1}{LT}\int_0^T\int_0^L  \mathbb{E}(\delta \mu_{ab}^2)dydt =\mu_0^2,
\end{equation}
where  $\mu_0^2$ is a new constant.

\section{Justification of expressions \eqref{e:deltamu_modes}. }\label{sec:appendix1}
Using the Hilbert-Schmidt projection of $\delta \mu$, we express the entries of the last  equation of system \eqref{eq:final_pb} as:
$$\begin{array}{c}
\partial_t\delta \mu(y,t)=\sum_{k=0}^{\infty}\partial_t\delta\mu_k(t)w_k(y)\\
D_{ab}\partial_{yy}\delta \mu(y,t)=-D_{ab}\sum_{k=0}^{\infty}\frac{k^2\pi^2}{L^2}\delta\mu_k(t)w_k(y)\\
\frac{2 \kappa  \chi_{\Sigma a}}{\alpha_{ab} \eta_f (1-\phi )}\partial_{yy}\delta P_f(y,t)=\sum_{k=0}^{\infty}\alpha_k\delta\mu_k(t)w_k(y)+\sum_{k=0}^{\infty}\tilde{\alpha}_{2k+1}\delta\mu_{2k+1}(t)\sinh\left(\frac{\tr{\Lambda}}{2}(L-2y) \right)=\\\sum_{k=0}^{\infty}\alpha_k\delta\mu_k(t)w_k(y)+\sum_{k=0}^{\infty}\sum_{l=0}^{\infty}\tilde{\alpha}_{2k+1}\beta_{2l+1}\delta\mu_{2k+1}(t)w_{2l+1}(y)
\end{array}$$
where, 
$$\alpha_k=-\frac{2 \pi ^2 k^2 }{\tau_a \left(\pi ^2 k^2+f^2 \right)}\text{, }\gamma_k=-\frac{4 \sqrt{2/L}  f^2  \text{sech}\left(f/2\right) }{\tau_a \left(\pi ^2 k^2+f^2 \right) \left(p f- f +2
     \tanh(f/2)\right)}\text{ and }\beta_k=\frac{2 \sqrt{2L}  f\cosh(f/2)}{ \pi ^2 k^2+f^2}.$$ 
We can therefore write two separate sets of linear first order ODE solving the even and odd modes of $\delta\mu$:
\begin{equation}\label{eq:modes}
\forall k\geq 0,\begin{array}{c}
\partial_t\delta\mu_{2k}(t)=d_{2k} \delta\mu_{2k}(t)+\frac{2}{\alpha_{ab}}\Gamma_{2k}(t)\\
\partial_t\delta\mu_{2k+1}(t)=d_{2k+1} \delta\mu_{2k}(t)+\beta_{2k+1}\sum_{l=0}^{\infty}\gamma_{2l+1}\delta\mu_{2l+1}+\frac{2}{\alpha_{ab}}\Gamma_{2k+1}(t),
\end{array}
\end{equation}
where 
$$d_k= -\frac{(\pi k)^2}{\tau_d}-\frac{2}{\tau_r}+\alpha_{k}.$$
To solve the system \eqref{eq:modes}, we consider it up to a finite mode $0\leq k\leq N$ and denote 
$\overline{\delta\mu}_E=(\delta\mu_{2k})_{k\geq 0}$ the vector of even modes of $\delta \mu$ and $\overline{\delta\mu}_O=(\delta\mu_{2k+1})_{k\geq 0}$ the vector of odd modes. More generally, the indices $E$ and $O$ will be used throughout the text to extract the even and odd components of a vector or a matrix.  We can thus rewrite \eqref{eq:modes} in matrix form as
\begin{equation}\label{eq:modes_mat}
\partial_t\overline{\delta\mu}_E=\mathbb{D}_E\overline{\delta\mu}_E+\frac{2}{\alpha_{ab}}\overline{\Gamma}_E\text{ and }\partial_t\overline{\delta\mu}_O=(\mathbb{D}_O+\mathbb{M}_O)\overline{\delta\mu}_O+\frac{2}{\alpha_{ab}}\overline{\Gamma}_O,
 \end{equation}
where $\mathbb{D}=\text{diag}(\overline{d})$ is the diagonal matrix $\mathbb{D}_{ij}=d_i\delta_{ij}$ and $\mathbb{M}=\overline{\beta}\overline{\gamma}^T$ is the matrix mixing the odd modes, $\mathbb{M}_{ij}=\beta_i\gamma_j$. We can express $\mathbb{D}=-(\pi^2/\tau_d) \bar{\mathbb{D}}-(2/\tau_r)\mathbb{I}-(2\pi^2/\tau_a)\bar{\mathbb{D}}(\pi^2\bar{\mathbb{D}}+f^2\mathbb{I})^{-1}$ where $\bar{\mathbb{D}}_{ij}=i^2\delta_{ij}$. 

The steady state solutions of \eqref{eq:modes_mat} are then given by,
\begin{equation}\label{eq:solutions_mu_modes}
\overline{\delta\mu}_E(t)=\frac{2}{\alpha_{ab}}\int_0^t\text{e}^{\mathbb{D}_E(t-u)}\overline{\Gamma}_E(u)du \text{ and }\overline{\delta\mu}_O(t)=\frac{2}{\alpha_{ab}}\int_0^t\text{e}^{(\mathbb{D}_O+\mathbb{M}_O)(t-u)}\overline{\Gamma}_O(u)du.
\end{equation}
For the even modes, as $\mathbb{D}_E$ is diagonal, it is straightforward to compute the integral as the exponential directly reads,
$$\forall\,t,\,(\text{e}^{\mathbb{D}_Et})_{ij}=\text{e}^{d_{2i}t}\delta_{ij}.$$
The computation of $\text{e}^{(\mathbb{D}_O+\mathbb{M}_O)t}$ requires more care. Since  $\mathbb{D}_O$ and $\mathbb{M}_O$ do not commute, we use the Trotter-Kato formula 
$$\text{e}^{(\mathbb{D}_O+\mathbb{M}_O)t}=\lim_{n\rightarrow \infty}\left( \text{e}^{\mathbb{D}_Ot/n}\text{e}^{\mathbb{M}_Ot/n}\right) ^n.$$
Next, as $\forall k\geq 1$, 
\begin{equation}\label{eq:power_property}
\mathbb{M}_O^k=(\overline{\beta}\overline{\gamma}^T)^k=s^{k-1}\mathbb{M}_O,
\end{equation}
where the scalar product $s$ reads $s=\overline{\beta}^T\overline{\gamma}$, we obtain,
$$\text{e}^{\mathbb{M}_Ot/n}=\sum_{k=0}^{\infty}\frac{(\mathbb{M}_Ot/n)^k}{k!}=\mathbb{I}+\sum_{k=1}^{\infty}\left( \frac{t}{n}\right)^k\frac{\mathbb{M}_O^k}{k!}=\mathbb{I}+\frac{\mathbb{M}_O}{s}\sum_{k=1}^{\infty}\left(\frac{st}{n}\right)^k\frac{1}{k!}=\mathbb{I}+\frac{\text{e}^{st/n}-1}{s}\mathbb{M}_O.$$
Injecting this expression in the Trotter-Kato formula we have
$$\left( \text{e}^{\mathbb{D}_Ot/n}\text{e}^{\mathbb{M}_Ot/n}\right) ^n=\text{e}^{\mathbb{D}_Ot}\left(\mathbb{I}+\frac{\text{e}^{st/n}-1}{s}\left( \text{e}^{-\mathbb{D}_Ot/n}\overline{\gamma}\right) \left(\text{e}^{\mathbb{D}_Ot/n}\overline{\beta} \right)^T  \right)^n,$$
which we expand with the binomial formula and a property similar to \eqref{eq:power_property} to reach
$$\left( \text{e}^{\mathbb{D}_Ot/n}\text{e}^{\mathbb{M}_Ot/n}\right) ^n=\text{e}^{\mathbb{D}_Ot}\left[ \mathbb{I}+\left( \text{e}^{-\mathbb{D}_Ot/n}\overline{\gamma}\right) \left(\text{e}^{\mathbb{D}_Ot/n}\overline{\beta} \right)^T\frac{\text{e}^{st/n}-1}{s\tilde{s}_n(t)}\sum_{k=1}^n\tbinom{n}{k} \tilde{s}_n(t)^k\right],$$
where,
$$\tilde{s}_n(t)=\frac{\text{e}^{st/n}-1}{s}\left(\text{e}^{\mathbb{D}_Ot/n}\overline{\beta} \right)^T\left( \text{e}^{-\mathbb{D}_Ot/n}\overline{\gamma}\right).$$
Thus,
$$\left( \text{e}^{\mathbb{D}_Ot/n}\text{e}^{\mathbb{M}_Ot/n}\right) ^n=\text{e}^{\mathbb{D}_Ot}\left[ \mathbb{I}+\frac{(\text{e}^{st/n}-1) (\left(1+\tilde{s}_n(t) \right)^n -1)}{s\tilde{s}_n(t)} \left( \text{e}^{-\mathbb{D}_Ot/n}\overline{\gamma}\right) \left(\text{e}^{\mathbb{D}_Ot/n}\overline{\beta} \right)^T\right]$$
and taking the limit when $n\rightarrow \infty$, we finally obtain
\begin{equation}\label{eq:exp_formula}
\forall\,t,\,\text{e}^{(\mathbb{D}_O+\mathbb{M}_O)t}=\text{e}^{\mathbb{D}_Ot}\left(\mathbb{I}+\frac{\text{e}^{st}-1}{s}\mathbb{M}_O \right).
\end{equation}
In the above formula, $s$ can be computed explicitly:
$$s=-\sum_{k=0}^{\infty}\frac{16  f^3   }{\tau_a \left(\pi ^2 (2k+1)^2+f^2 \right)^2 \left(p f- f +2
     \tanh(f/2)\right)}=\frac{(f-\sinh (f)) \text{sech}^2\left(f/2\right)}{\tau_a \left(f p-f+2 \tanh \left(f/2\right)\right)}.$$
In index notation, formula \eqref{eq:solutions_mu_modes} therefore leads to the expressions given in the main paper:
$$
\forall\,i\geq 0,\left\lbrace \,\begin{array}{c}
\delta\mu_{2i}(t)=\frac{2}{\alpha_{ab}}\int_0^t\text{e}^{d_{2i}(t-u)}\Gamma_{2i}(u)du \\
\delta\mu_{2i+1}(t)=\frac{2}{\alpha_{ab}}\int_0^t\text{e}^{d_{2i+1}(t-u)}\Gamma_{2i+1}(u)du+\frac{2}{\alpha_{ab}s}\int_0^t\text{e}^{d_{2i+1}(t-u)} \left( \text{e}^{s(t-u)}-1\right)  \beta_{2i+1}\sum_{k=0}^{\infty}\gamma_{2k+1}\Gamma_{2k+1}(u)du.
\end{array}\right.
$$

\renewcommand\thefigure{\arabic{figure}}


\end{document}